\documentclass[12pt]{iopart}

\usepackage{iopams}
\usepackage{graphicx}
\begin{document}

\topical[Naber, Faez and Van der Wiel -- Organic
Spintronics]{Organic Spintronics}

\author{W.J.M. Naber, S. Faez and W.G. van der Wiel}

\address{SRO NanoElectronics and NanoElectronics group, MESA$^+$ Institute for Nanotechnology,
University of Twente, PO Box 217, 7500 AE Enschede, The Netherlands}

\begin{abstract}
In this paper we review the recent field of organic spintronics,
where organic materials are applied as a medium to transport and
control spin-polarized signals. The contacts for injecting and
detecting spins are formed by metals, oxides, or inorganic
semiconductors. First, the basic concepts of spintronics and
organic electronics are addressed and phenomena which are in
particular relevant for organic spintronics are highlighted.
Experiments using different organic materials, including carbon
nanotubes, organic thin films, self-assembled monolayers and
single molecules are then reviewed. Observed magnetoresistance
points toward successful spin injection and detection, but
spurious magnetoresitance effects can easily be confused with spin
accumulation. A few studies report long spin relaxation times and
lengths, which forms a promising basis for further research. We
conclude with discussing outstanding questions and problems.
\end{abstract}

\maketitle

\section*{Introduction}
\label{intro}


Organic spintronics is a new and promising research field where
organic materials(\footnote{ The word `organic' stems from the
19th-century belief that certain compounds, termed organic
materials, could only be formed in living organisms. This belief
turned out to be incorrect, but the definition is still somewhat
ambiguous. Organic materials are now often defined as those
materials which contain carbon-hydrogen bonds. This definition would
exclude fullerenes like carbon nanotubes, as they consist of C only.
Fullerenes are however mostly considered organic materials, as we
also do in this review.}) are used to mediate or control a spin
polarized signal. It is hence a fusion of organic electronics
\cite{Farchioni,Cuniberti,Klauk} and spin electronics (or
\textit{spintronics})
\cite{Prinz1,Prinz2,Wolf,DasSarma,Gregg,Zutic}. Organic materials,
on the one hand, open the way to cheap, low-weight, mechanically
flexible, chemically interactive, and bottom-up fabricated
electronics. The application of the electron's spin (instead of or
in addition to its charge), on the other hand, allows for
non-volatile devices. Spintronic devices are also potentially faster
and consume less electrical power, since the relevant energy scale
for spin dynamics is considerably smaller than that for manipulating
charges.

\begin{figure}[htbp]
\center{  \includegraphics[width=4in]{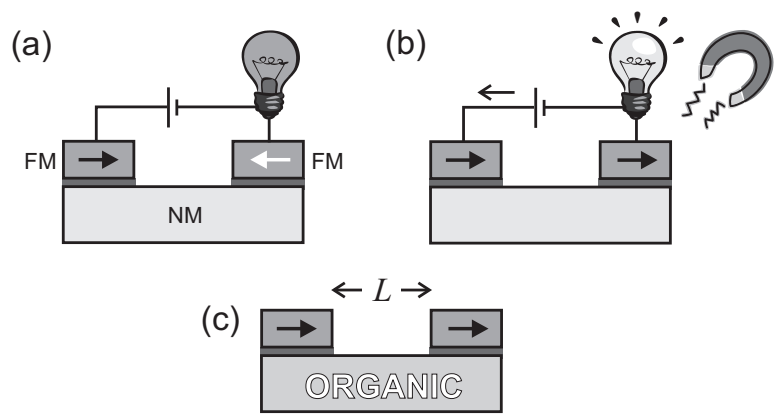}}
  \caption{Schematic representation of a spin valve. Two ferromagnetic (FM) contacts
  (magnetization denoted by arrows) are separated by a non-magnetic (NM) spacer
  (bottom). One of the contacts is used as spin injector, the other one as
  spin detector. A tunnel barrier in between the FM contact and the NM spacer
  can enhance the spin signal. The light bulb schematically indicates (a) low
  conductance in the case of anti-parallel magnetization, and (b) large conductance
  for parallel magnetization. (c) Spin valve with organic spacer.}
  \label{schemspin}
\end{figure}

Figure \ref{schemspin}(a) schematically shows the canonical example
of a spintronic device, the \emph{spin valve} (SV). Two
ferromagnetic (FM) contacts with different coercive fields ($H_c$),
applied as spin injector and spin detector, respectively, are
separated by a non-magnetic (NM) spacer. The role of the spacer is
to decouple the FM electrodes, while allowing spin transport from
one contact to the other. The electrical resistance depends on the
relative orientation of the magnetization of the two FM contacts.
The relative orientation can be tuned by an external magnetic field
between the anti-parallel (AP) and parallel configuration (P), as in
Figure \ref{schemspin}(b). As discussed later, the resistance is
usually higher for the AP configuration, an effect referred to as
giant magnetoresistance (GMR)(\footnote{The qualification `giant' is
used to distinguish the effect from anisotropic magnetoresistance
(AMR). AMR refers to the dependence of the electrical resistance on
the angle between the direction of the electrical current and the
orientation of the magnetic field \cite{Thomson}. The observed GMR
effects are about 2 orders of magnitude larger.}). The spacer
usually consists of a NM \emph{metal}, or a thin insulating layer
(in the case of a magnetic tunnel junction, MTJ). The
magnetoresistance (MR) effect in the latter case is referred to as
tunnel magnetoresistance (TMR). Only very recently, Lou and
coworkers have demonstrated all-electric spin injection and
detection using the inorganic \emph{semiconductor} GaAs as NM spacer
\cite{Lou}.

In an organic spintronic device, the NM spacer consists of an
organic material, see figure \ref{schemspin}(c). The device of
figure \ref{schemspin}(c) is actually a \emph{hybrid} device, since
inorganic (FM contacts) and organic (NM spacer) materials are
combined. In principle, also the FM contacts could be made out of
organic materials (i$.$e$.$ organic ferromagnets), resulting in an
all-organic spintronic device. Although organic materials with FM
properties do exist \cite{orgmagnet1,orgmagnet2,orgmagnet3}, to the
best of our knowledge all-organic spintronic devices have not been
realized so far. This review therefore focuses on structures with
the hybrid geometry of figure \ref{schemspin}(c).

The field of organic spintronics not only combines the
aforementioned advantages of organic electronics and spintronics, it
has particularly attracted attention because of the potentially very
long spin relaxation times in organic materials \cite{Sanvito}.
Using electron paramagnetic resonance (EPR) measurements,
room-temperature spin relaxation times in the range $10^{-7} -
10^{-5}$ s have been found \cite{Krinichnyi} (as compared to
$\sim$$10^{-10}$ s in metals \cite{Jedema}).

The spin relaxation time, $\tau_s$, or spin lifetime, is given by
\begin{equation}
\frac{1}{\tau_s} = \frac{1}{\tau_{\uparrow \downarrow}} +
\frac{1}{\tau_{\downarrow \uparrow}} \label{reltime}
\end{equation}
with the spin flip time $\tau_{\uparrow \downarrow}$ indicating the
average time for an up-spin to flip to a down-spin, and
$\tau_{\downarrow \uparrow}$ for the reverse process. The spin
relaxation time is a key parameter in spintronic devices, as it sets
the timescale -- and hence the length scale -- for loss of spin
polarization. The spin relaxation length, $l_s$, is related to the
spin relaxation time as
\begin{equation}
l_s = \sqrt{\frac{\tau_s}{4 e^2 N(E_F) \rho_N}}
 \label{rellength1}
\end{equation}
in the case of a NM metal or a degenerate Fermi gas semiconductor
\cite{Fert2,Fert3}. Here $N(E_F)$ is the density of states (DOS) at
the Fermi level, and $\rho_N$ the resistivity of NM spacer material.
For a semiconductor in the non-degenerate regime, $l_s$ is given by
\cite{Fert2,Fert3}
\begin{equation}
l_s = \sqrt{\frac{k_B T \tau_s}{2 n e^2 \rho_N}},
 \label{rellength2}
\end{equation}
where $k_B$ is the Boltzmann constant, $T$ the temperature, and $n$
the total number of carriers.

For the SV device of figure \ref{schemspin} to work properly, the
distance $L$ between the FM contacts should be smaller than the spin
relaxation length: $L \ll l_s$. In inorganic materials, the dominant
spin relaxation mechanisms are spin-orbit coupling and hyperfine
interaction, which both turn out to be weak in organic materials, as
discussed in section \ref{spinrelax}. The geometry of figure
\ref{schemspin} can be used to determine $l_s$ by varying the
contact distance $L$, and measuring the decay of the MR signal (see
section \ref{CPP GMR}. Such an all-electric determination of $l_s$
is particularly interesting for organic conductors. The small
spin-orbit coupling results in the absence of optical selection
rules that are taken advantage of in spin relaxation measurements in
(inorganic) II-VI and III-V semiconductors
\cite{Kikkawa97,Kikkawa98,Kikkawa99,Malajovich}. Note that (AP) FM
contacts to organic light-emitting diodes (OLEDs) have been proposed
to increase their emission efficiency, by increasing the relative
amount of singlet excitons \cite{Arisi,Bergenti,Hayer}. Injection of
spin-polarized carriers from other FM elements like gadolinium,
which is not a transition metal, into organic semiconductors has
also been investigated in functional OLEDs in order to generate
magnetic field dependent luminescence \cite{2003-davis-JAP}.

For the lateral GMR geometry of figure \ref{schemspin} it is
essential that the injected spin current can be transferred over a
length $L$ with a minimum of spin relaxation. Besides the spin
relaxation time, the conductivity of the organic conductor needs
therefore to be sufficiently large. Whereas the long spin relaxation
time is a clear advantage of organic materials, the relatively low
conductivity of most organic conductors is a serious point of
consideration. However, important progress has been made in recent
years, see section \ref{orgelec}. Note that in organic TMR devices
the organic spacer forms a tunnel barrier, where the organic
material obviously should be insulating (see section \ref{TMR}).

Another important issue in organic electronics in general, and in
organic spintronics in particular, is \emph{contacting} the organic
material. Organic materials are usually fragile and the standard
microfabrication techniques used for contacting inorganic materials,
often introduce considerable damage, making the contacts poorly
defined. As spin injection and detection occurs at the interface of
the FM contacts and the organic material, the quality of this
interface is of crucial importance.

In this Review, we present an overview of the experiments in the
field of organic spintronics so far. As the field is still
relatively young and rapidly expanding, this Review cannot (and is
not intended to) be the `final word' on organic spintronics. Instead
it is meant as a comprehensive reference for those who like to
explore this new area of research. In the first part of the Review,
we briefly discuss the field of organic electronics (section
\ref{orgelec}), key spintronic concepts (section \ref{spintronics}),
and spin relaxation (section \ref{spinrelax}). Special attention is
given to spurious MR effects that can obscure the desired spintronic
characteristics, and are therefore important for the correct
interpretation of experimental results. In the second part,
experiments on organic spintronic devices are discussed. We start
with carbon nanotube experiments (section \ref{CNT exp}), followed
by experiments on organic thin films of small molecules and
polymers, and self-assembled monolayers (section \ref{OTF exp}). We
conclude in section \ref{Concl}.

\section*{PART I}

\section{Organic electronics}
\label{orgelec}

Organic materials were for long time only associated with electrical
insulators. In the last century, however, the idea of organic
electronics arose. On the one hand, there was the wish to use
organic materials as (semi-)conductors in bulk or thin film. On the
other hand, the concept was put forward to use single molecules as
electrical components, such as switches and diodes. The latter field
is often referred to as \emph{single-molecule electronics} or
\emph{molecular electronics} \cite{Petty}. The advantages of organic
materials include chemical tuning of electronic functionality, easy
structural modifications, ability of self-assembly and mechanical
flexibility. These characteristics are exploited for large-area and
low-cost electronic applications
\cite{Voss,Reese,Forrest,Boer_review,polymervision}. Single
molecules may eventually form the ultimately miniaturized
electronics \cite{Cuniberti}. In this section, we briefly discuss
the main developments in organic electronics.

\subsection{Organic thin films}
\label{OTF}

Present-day electronics is dominated by the Si/SiO$_2$ metal-oxide
semiconductor field-effect transistor (MOSFET), where a gate voltage
forms an inversion layer in between the source and drain contacts of
the transistor \cite{Sze}. The ability to drastically change the
carrier density in semiconductors by doping and electrical gating is
essential in electronics.

Driven by the technological potential of organic materials, interest
arose in organic semiconductors. Present-day organic semiconductors
are mainly $\pi$-conjugated materials, usually divided in
\emph{polymers} and \emph{small molecules}, with $\sim$1.5-3.5 eV
band gaps \cite{Ruden04}. Thin, amorphous or poly-crystalline films
of these materials have been successfully applied in organic
light-emitting diodes (OLEDs) \cite{Tang87,Friend}, photovoltaic
cells \cite{Brabec,Forrest03}, and field-effect transistors (FETs)
\cite{Sirringhaus99,Dimitrakopoulos}. Thin-film technology does not
require high temperatures and lattice matching as in the case of
inorganic heterostructures. Significant improvement in the
performance of those devices was realized in the last few years.

\begin{figure}[htbp]
\center{  \includegraphics[width=7cm]{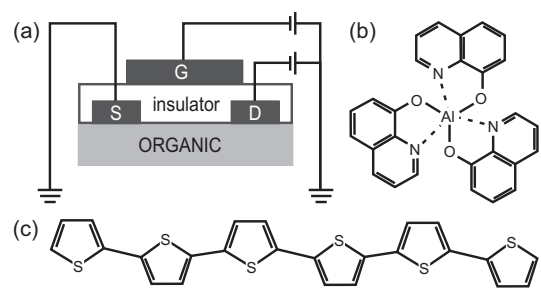}}
  \caption{(a) Schematic layout of an organic thin-film transistor.
  The gate (G) electrode, which induces a conducting channel, is
  separated from the organic film by an insulator. The current
  through the organic material is injected and collected by source
  (S) and drain (D) contacts. Structures of two $\pi$-conjugated
  molecules are given: (b) 8-hydroxy-quinoline (Alq$_3$) and (c) the oligomer $\alpha$-sexithienyl (T$_6$).}
  \label{OTFT}
\end{figure}

Control of the carrier density by doping, as is done in inorganic
(extrinsic) semiconductors, is not straight-forward for most organic
semiconductors as they are not pure enough. The effect of doping
only manifests itself at high doping levels, where the behaviour is
more metallic than semiconducting. Therefore, in organic transistors
the thin-film-transistor (TFT) geometry [see figure \ref{OTFT}(a)]
is used rather than that of the MOSFET. In an organic TFT, a
conducting channel is capacitively induced at the interface between
the dielectric and the organic material. The charge does thus not
originate from dopants as in MOSFETs. Carriers are injected into the
conducting channel from metallic contacts. Electrical conduction in
(disordered) thin films normally results from carrier hopping
between localized states (see section \ref{transport}), and not from
band-like transport through delocalized states, as typical for
inorganic semiconductors.

\subsubsection{Polymers}
\label{polymers}

Research on organic semiconductors first focused on improving the
conductance of organic polymers. In 1963 high conductivity was
reported in iodine-doped and oxidized polypyrrole \cite{Bolto}.
Research on organic conductors was further boosted by the discovery
of high conductivity in oxidized, iodine-doped polyacetylene
\cite{Chiang,Shirakawa}, for which Heeger, MacDiarmid and Shirakawa
received the Nobel Prize in Chemistry in 2000. `High conductivity'
is relative in this respect, as almost all known conductive polymers
are semiconductors with a low electronic mobility(\footnote{Under
certain circumstances polymers can actually become metallic and even
superconducting.}). The maximum mobilities of polymer films are
typically 0.1 cm$^2$(Vs)$^{-1}$ \cite{Klauk}. The big advantage of
polymer films though is that there are well-developed deposition
techniques available to process them.

In polymer (or plastic) electronics, especially the conjugated
polymers are important \cite{Sirringhaus}. These are polymers in
which a sequence of alternating single and double bonds is present
in the polymer chains. The wave function of one of the four
electrons of carbon, which forms a $\pi$-bond with its neighboring
electrons, is in this case delocalized along the polymer and its
mobility along one polymer can be rather high \cite{Boer2}. Next to
the conduction within one molecule, also the interaction of a
$\pi$-system with the $\pi$-system of a neighboring molecule
determines the conductivity of the polymer film. The Peierls
instability \cite{Peierls} causes that in practice all conjugated
polymers act as semiconductors. The structure of polymer films is
rather irregular (more or less `spaghetti-like'), strongly limiting
their mobility.

\subsubsection{Small molecules}
\label{smallmolecules}

More ordered films can be realized with small molecules, resulting
in higher mobilities ($\sim$1 cm$^2$(Vs)$^{-1}$). One of the
materials most commonly used for (p-type) OTFTs is pentacene with a
highest reported mobility of 6 cm$^2$(Vs)$^{-1}$ \cite{Wang}. Most
thin films of small molecules are grown by vapor deposition. The
film-dielectric interface turns out to be of great importance for
the performance of the OTFT and a lot of effort has been put in
improving this interface, e$.$g$.$ by introducing self-assembled
monolayers \cite{Campbell}. The small organic molecule Alq$_3$ and
the oligomer T$_6$, examples of organic materials that have been
applied in spintronic devices, are shown in figure \ref{OTFT}(a) and
(b), respectively.

\subsection{Single-crystals}
\label{osc}

Single-crystals of organic semiconductors \cite{Silinsh} like
rubrene and pentacene, are similar to the single-crystal structures
of inorganic electronics. Ultra-pure organic single-crystals (OSCs)
can be grown and their electronic properties are well-reproducible
\cite{Boer_review}. In OSCs grain boundaries are eliminated and the
concentration of charge traps is minimized \cite{Sundar}, making
them suitable for studying the intrinsic electronic properties of
organic materials and the physical limitations of organic FETs. In
contrast, thin films of polymers or small molecules are often
strongly affected by imperfections of the film structure and by
insufficient purity of the materials \cite{Brocks}. The electric
mobilities increased largely recently, reaching room-temperature
values of 35 cm$^2$(Vs)$^{-1}$ in pentacene \cite{Jurchescu} and 20
cm$^2$(Vs)$^{-1}$ in rubrene \cite{Podzorov}. Single-crystals cannot
be deposited from solution, but instead the physical vapour
transport (PVT) method is used \cite{Klauk,Laudise}. The techniques
for fabricating OTFTs with as-grown OSCs has been reviewed in
\cite{Boer_review}. Recently, selective growth of OSCs on domains of
octadcyltriethoxysilane was reported \cite{Briseno06}.

\subsection{Single-molecule electronics}
\label{singlemolecules}

In a 1974 paper, Aviram and Ratner \cite{Aviram} introduced the
concept of a molecular rectifier, based on the idea of
`donor-acceptor' systems already put forward in the 1940s by
Mulliken and Szent-Gy\"{o}rgi \cite{Szent41}. The first experimental
study of single-molecule conductance was reported by Reed \textit{et
al}$.$ in 1997 \cite{Reed}. One of the most important issues in
single-molecule electronics is the contacting of the molecule with
(metal) electrodes \cite{Dadosh}. Obviously the contact spacing
needs to be very small, typically on the order of 1 nm. The nature
of the molecule-metal interface is of crucial importance to the
transport properties \cite{Hipps}. Having good mechanical contact
does not automatically imply good electrical contact. End-group
engineering offers the possibility to chemically anchor the
molecules to metal contacts. Apart from hooking up a single molecule
to source and drain contacts, effective gating of the molecule is
rather difficult due to screening of the nearby metallic contacts.
Many different nano-contacting schemes have been developed over the
last decade. Examples include mechanical break junctions
\cite{Reed}, nanopores \cite{Chen}, electromigration \cite{Park} and
conducting-probe atomic force microscopy \cite{Cui}.

\subsection{Carbon nanotubes}
\label{CNT}

Carbon nanotubes (CNTs) are carbon cylinders of a few nanometers in
diameter and up to several millimeters in length
\cite{1999-dekker-phystoday,CNTbook,Baughman}. They were discovered
by Sumio Iijima in 1991 \cite{Iijima}. CNTs belong to the fullerene
structural family, which for example also includes the C$_{60}$
`buckyball' molecule. They can be thought of as rolled up graphene
sheets capped at their ends with hemispheres of the buckyball
structure. Single-walled carbon nanotubes (SWCNTs) consist of a
single carbon cylinder, whereas multi-walled carbon nanotubes
(MWCNTs) are made up of multiple concentric cylinders.

The electrical properties of a SWCNT are determined by the way the
graphene sheet is rolled up, expressed by the chiral vector $(n,m)$,
where the integers $n$ and $m$ denote the number of unit vectors
along two directions in the honeycomb crystal lattice of graphene
\cite{1999-dekker-phystoday}. If $2n + m = 3q$ (with $q$ an
integer), the SWCNT is metallic, otherwise semiconducting. Metallic
nanotubes can have an electrical current density of 1 TA/cm$^3$
(i$.$e$.$ $\sim$1,000 times larger than that of metals such as
silver and copper) \cite{saito}. The current in MWCNTs is usually
thought to mainly flow through the outermost shell
\cite{1998-Frank-Science,1999-Schonenberger-APA}.

CNTs have attracted a lot of interest because of their exceptional
electronic and mechanical properties \cite{1999-dekker-phystoday}.
CNTs have been applied as FETs in logic circuits
\cite{2001-dekker-science}, and have been been widely proposed for
organic electronics applications
\cite{carbon_electronics,carbon_electronics2,carbon_electronics3,carbon_electronics4}.
More recently, also spin injection and transport in CNTs is
intensely studied. The combination of high charge mobility,
negligible spin-orbit coupling (light C atoms) and weak hyperfine
interaction(\footnote{$^{12}$C has nuclear spin zero, but the
isotope $^{13}$C (1.1\% abundancy) has nuclear spin 1/2. The
concentration $^{13}$C can however be reduced by isotopic
purification.}) holds the promise of very long spin relaxation
lengths. The first organic spintronic device was reported by
Tsukagoshi, Alphenaar and Ago, and consisted of a MWCNT contacted
by Co contacts \cite{1999-Tsukagoshi-Nature}.

SWCNTs have also been put forward as ideal 1D electronic systems in
which Tomonaga-Luttinger-liquid (TLL) behaviour should be
observable. A TLL is a model for interacting electrons (or other
fermions) in a 1D conductor, where the conventional Fermi liquid
model breaks down \cite{Tomonaga,Luttinger}. The elementary
excitations of the TLL are formed by separate charge and spin waves,
in contrast to the quasiparticles in a Fermi liquid, which carry
both spin and charge. The property that the charge and spin waves
are mutually independent in a TLL is referred to as
\emph{spin-charge separation}. Spin-polarized transport in CNTs
could shine more light on the electron-electron interactions in 1D
systems. Balents and Egger theoretically showed that the spin-charge
separation in a TLL modifies spin transport \cite{Balents}.
Tunnelling into a TLL is suppressed due to the strong e-e
interactions in a 1D electronic system. Typical of a TLL is a power
law behaviour: $dI/dV \sim V^{\alpha}$.

The small dimensions of CNTs allows for the definition of a quantum
dot (QD) inside the CNT. In this way, one can study the interplay of
spin transport with single-charging and quantum confinement effects.
In a QD also a \emph{single} electron spin can be confined and
manipulated \cite{Tokura06}. This is particularly interesting for
realizing single-spin quantum bits for quantum computing and quantum
information \cite{LossDiVincenzo}.

\subsection{Electronic transport in organic materials}
\label{transport}

\subsubsection{Hopping vs band transport}

Charge injection and transport in organic materials are still not
understood in full detail. In general, one can distinguish two main
charge transport mechanisms: \emph{hopping} and \emph{band
transport}. The hopping mechanism is typical for disordered
materials such as the organic thin films of section \ref{OTF}.
Transport occurs via hopping between localized molecular states
\cite{Zuppiroli} and strongly depends on parameters like
temperature, electric field, traps present in the material and the
carrier concentration
\cite{Boer_review,Vissenberg,Nelson,Coehoorn,Pasveer}. This leads to
a much smaller mobility than via delocalized band states, as in
crystalline inorganic semiconductors \cite{Coehoorn}. Band-like
conduction in organic materials is only expected at low temperature
for highly ordered systems \cite{Hannewald,Horowitz}, such as the
OSCs of section \ref{osc}, when the carrier mean free path exceeds
the intermolecular distance \cite{Silinsh}. The valence band then
generally originates from the overlap of the HOMO levels, and the
conduction band from the overlap of the LUMO levels of the molecules
\cite{Farchioni}.

\subsubsection{p-type and n-type conduction}

It should be noted that the terms `n-type' and `p-type' in organic
semiconductors do not have the same meaning as for inorganic
semiconductors. In the inorganic case, `n-type' (`p-type') refers to
doping with electron donors (acceptors). In the organic case
however, an `n-type' (`p-type') material is a material in which
electrons (holes) are more easily injected than holes (electrons)
\cite{Klauk}. In organic semiconductors, p-type conduction is much
more common than n-type conduction, i$.$e$.$ in most organic
materials hole transport is favored. This has been explained by the
fact that electrons are much more easily trapped at the
organic-dielectric interface than holes \cite{Dimitrakopoulos,Chua}.
There are a few reports on n-type organic semiconductors
\cite{Jarrett,Kobayashi,Katz,Malenfant}, and also \emph{ambipolar}
organic materials (showing both p-type and n-type behaviour,
dependent on the gate voltage) \cite{Chua,Boer4} have been
identified. However, the electron mobility is generally considerably
lower than the hole mobility. For electronic logic it would be
favorable to combine n- and p-type organic materials to realize
complementary circuitry (as in CMOS technology \cite{Sze}).

\subsubsection{Polarons}

As the intermolecular (van der Waals) forces in organic materials
are much weaker than the covalent and ionic bonds of inorganic
crystals, organic materials are less rigid than inorganic substances
\cite{Pope}. A propagating charge carrier is therefore able to
locally distort its host material. The charge carrier combined with
the accompanying deformation can be treated as a quasi-particle
called a \emph{polaron} \cite{Brazovskii}. A polaron carries spin
half, whereas two near-by polarons (referred to as a
\emph{bipolaron}) are spinless \cite{Chance}. Polaron formation
generally reduces the carrier mobility
\cite{Hannewald,Johansson,Fu,Wei1,Wei2,Stoneham,Bussac}.
It is more and more realized that electronic transport in organic
materials is not only determined by the characteristics of the
organic conductor itself, but also by the interplay with the
adjacent dielectric layer \cite{Stassen,Hulea}. It is therefore
important to find a suitable conductor-dielectric combination
\cite{Chua}.

\subsubsection{Contacting}

Apart from the conduction mechanism, also the charge injection into
the organic material is of crucial importance to the performance of
the device. The charge injection mechanism strongly depends on the
interface between the contact and organic material. This can involve
impurities, structural defects, charging, dangling bonds, dipoles,
chemical moieties and other effects, in which also the fabrication
method of the device plays a significant role.

Carrier injection across the metal-organic interface is determined
by the energy barrier height and the density of states (DOS) at the
Fermi level ($E_F$) of the metal contact
\cite{Davids97,CampbellScott}. Contact resistance can be the result
of a mismatch of the HOMO (for p-type semiconductors) or LUMO (for
n-type semiconductors) with respect to the work function of the
electrode metal. The resulting Schottky barrier gives rise to
non-linear (diode-like) behaviour. The interface resistance depends
exponentially on the barrier height, and linearly on the DOS of the
metal contact at $E_F$.

The Schottky barrier at the interface between a metal and organic
semiconductor usually directly scales with the metal work function,
as opposed to the case of inorganic semiconductors, where the
Schottky barrier only weakly depends on the metal work function
\cite{Ruden04,Campbell01}. Hence, low-work-function metals such as
Ca are used to inject electrons, and high-work-function metals such
as Au or InSnO (ITO) are used to inject holes into an organic
semiconductor.

Since organic materials in general are rather fragile, conventional
contacting methods can easily damage the material, causing a bad
interface between the material and the electrode. A number of
techniques have been developed for non-destructively contacting,
including soft lithography (e.g. micro transfer printing)
\cite{Xia,Hur,Hsu}, ink-jet printing \cite{Sirringhaus2},
solution-based methods \cite{Sirringhaus,Ong} and vapour phase
deposition \cite{Reese,Forrest}. The interface properties are
especially important for spin injection, as is discussed in more
detail in section \ref{mismatch}.

\subsubsection{Single-molecule transport}

Transport through a single molecule is very different from bulk
transport. At sufficiently low temperatures transport can be
dominated by Coulomb blockade and quantum confinement effects
\cite{LeoNATO,Wiel,Fujisawa,Naber}. In the simplest model only
transport through one molecular level is considered. When this level
is between the Fermi levels of the two leads, current will flow
\cite{Datta}. A more accurate method which is by far most used is
the non-equilibrium Green's function (NEGF) method \cite{Haug}.

A number of methods has been developed for calculating transport
\cite{Rocha2,Taylor,Xue,Brandbyge,Palacios,Pecchia,Bratkovsky}. Some
of these (\cite{Rocha2,Taylor,Xue,Palacios}) are also applicable for
spin-polarized transport, e$.$g$.$ in molecular SVs, consisting of a
molecule sandwiched between two nanoscale FM contacts \cite{Strunk}.
Rocha \textit{et al}$.$ \cite{Rocha} show that it is possible to
obtain a very high spin-dependent signal. They used the code SMEAGOL
\cite{Rocha2} (spin and molecular electronics in an atomically
generated orbital landscape). This code combines the NEGF method
with the density-functional-theory code SIESTA (Spanish Initiative
for Electronic Simulations with Thousands of Atoms) \cite{Soler}.
The code SMEAGOL is especially designed for spin-polarized
transport.

Emberly and Kirczenow \cite{Emberly} have theoretically reproduced
experiments on a gold break junction bridged with benzenedithiol
molecules with a semi-empirical model. They extend this model to
break junctions formed by nickel, and systems with a nickel STM tip
scanning a nickel substrate covered with a bezenethiol monolayer. In
both cases they find spin-valve behaviour with this model.

\section{Spintronic concepts}
\label{spintronics}

In this section we briefly discuss the physical mechanisms of TMR,
GMR, and the conductivity mismatch problem. These concepts have been
originally developed and studied for inorganic systems, but are also
crucial for designing and understanding organic spintronic devices.
We also discuss a number of `spurious' MR effects that can easily be
mistaken for the desired spin valve effect.

\subsection{Historical perspective}
\label{spintronics history}

If a material or device changes its electrical resistance under the
influence of an external magnetic field, this property is generally
referred to as \emph{magnetoresistance}. The first known phenomenon
where the electrical resistance is altered by the direction of a
magnetic moment is called anisotropic magnetoresistance (AMR),
discovered in 1857 by Thomson \cite{Thomson}. AMR originates from
the larger probability for s-d scattering of electrons in the
direction of the magnetic field. The electrical resistance is
maximum for current and magnetic field parallel.

In 1973, Tedrow and Meservey determined for the first time
experimentally the spin polarization of the conduction band in a FM
material, using a FM/tunnel barrier/superconductor junction
\cite{Tedrow}. This work led to the discovery of TMR in FM/tunnel
barrier/FM junctions by Julli\`{e}re in 1975 \cite{Julliere}. In TMR
structures the tunnelling current is proportional to the product of
the DOS for each spin subband, and is hence dependent on the
relative orientation of the magnetizations in both FM layers. As
this relative orientation depends on the magnetic history, a TMR
structure can be used as a memory \cite{Moodera, Filip}. TMR is
therefore a pure \emph{interface effect} and does not require spin
transport in the NM layer \cite{Filip}.

With the discovery of GMR in 1988, for the first time spin-polarized
\emph{transport} through a NM metal was demonstrated. GMR was
discovered independently by Fert \emph{et al$.$} \cite{Baibich} and
Gr\"{u}nberg \textit{et al}$.$ \cite{Binasch}, and triggered a
tremendous amount of research on spintronic devices. The underlying
mechanisms of GMR differ fundamentally from that of TMR and are
discussed in more detail in section \ref{GMR}. The field of
spintronics was very much stimulated by the commercial success of
GMR devices. IBM already produced the first GMR-based harddisk read
head in 1997 \cite{Wallstreet}.

One of the long-standing goals in the spintronics community is the
realization of an active device that combines electric control of
the source drain current as in transistors with the memory effect of
spin valves. In 1990 Datta and Das proposed a FET device based on a
2D electron gas with FM contacts \cite{Datta2}. In the Datta-Das
`spin-FET', the current modulation between the FM contacts arises
from spin precession induced by gate-controllable spin-orbit
interaction. However, it was found that a strictly 1D-ballistic
channel is required for this purpose \cite{Tang}.

The wish to combine semiconductor and spintronic concepts,
stimulated efforts to inject spins into a semiconductor. Using an
all-optical pump-and-probe technique, Kikkawa {\it et al.} succeeded
in injecting spins in II-VI and III-V semiconductors and measuring
the spin (ensemble) coherence time \cite{Kikkawa97}. Very long
coherence times up to 1 $\mu$s have been measured in GaAs
\cite{Kikkawa98,Kikkawa99,Malajovich}. The first \emph{electrical}
injection into a semiconductor was demonstrated by Fiederling
\textit{et al}$.$ \cite{Fiederling}, although the spin detection is
still optical in this case. As mentioned before, only very recently
an all-electrical spin injection and detection scheme was
demonstrated for an inorganic semiconductor \cite{Lou}.

One of the major obstacles for spin injection/detection in
semiconductor devices is the so-called conductivity mismatch between
the semiconductor spacer and the metallic FM contacts. This issue is
addressed in section \ref{mismatch}.

\subsection{Tunnel magnetoresistance}
\label{TMR}

TMR originates from the difference in the DOS at $E_F$ between
spin-up $N_{\uparrow}(E_F)$ and spin-down $N_{\downarrow}(E_F)$
electrons. Given the conservation of spin orientation during
tunnelling, electrons can only tunnel from a given spin subband in
the first FM contact to the \emph{same} spin subband in the second
FM contact, as schematically depicted in figure \ref{TMR}. The
tunnel rate is proportional to the product of the corresponding spin
subband DOSs at $E_F$, and hence on the relative magnetization
orientation of the contacts. Consequently, the resistance in the P
configuration [figure \ref{TMR}(b)] is lower than in the AP
configuration [figure \ref{TMR}(c)].

\begin{figure}[htbp]
\center{  \includegraphics[width=4cm]{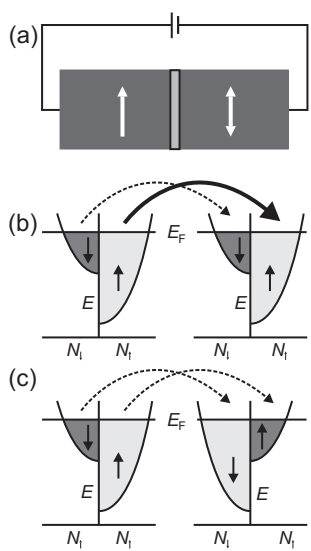}}
  \caption{(a) Schematic representation of a TMR device, consisting
  of two FM materials (dark gray) separated by a tunnel barrier (light gray). The
  magnetization
  can be parallel (P) or anti-parallel (AP), denoted by the arrows.
  Spin subbands of the FM materials are given for the P (b)
  magnetization and AP magnetization (c). The dashed (solid)
  arrow represents low (high) spin current.}
  \label{TMR}
\end{figure}

Based on the work of Tedrow and Meservey \cite{Tedrow}, assuming
spin and energy conservation, Julli\`{e}re derived a compact
expression for the difference in resistance between the P and AP
configurations, the TMR ratio(\footnote{Note that the following,
alternative, definitions of the TMR ratio are also frequently used:\\
$TMR' \equiv \frac{R_{AP}-R_{P}}{R_{AP}} = \frac{2P_{1}P_{2}}{1 +
P_{1}P_{2}}$, and $TMR'' \equiv  2 \frac{R_{AP}-R_{P}}{R_{AP}+R_{P}}
= 2P_{1}P_{2}$.})
\begin{equation}
TMR \equiv \frac{R_{AP}-R_{P}}{R_{P}} = \frac{G_P - G_{AP}}{G_{AP}}
= \frac{2P_{1}P_{2}}{1 - P_{1}P_{2}},
\label{Julliere}
\end{equation}
where $R_{P(AP)}$ is the resistance in the P (AP) configuration,
G$_{P(AP)}$ the conductance for the P (AP) configuration, and
$P_{1(2)}$ the polarization of the first (second) FM contact with
\begin{equation}
P_i = \frac{N_{i \uparrow}(E_F)-N_{i \downarrow}(E_F)}{N_{i
\uparrow}(E_F)+N_{i \downarrow}(E_F)}, \hspace{1cm} i=1,2.
\end{equation}
Although the Julli\`{e}re model gives a good basic insight, it
cannot explain a number of experimental observations like the
dependence on temperature and bias voltage, the material the tunnel
barrier is made of, and the height and width of the barrier
\cite{Moodera}. A model incorporating all these effects is still
lacking.

The Julli\`{e}re model treats the FM contacts as independent, and is
only valid for a square barrier. In real devices, the carrier wave
functions from both FM contacts overlap, and a finite bias voltage
gives a non-square barrier shape. Slonczewski \cite{Slonczewski}
altered the Julli\`{e}re-model, taking into account the permeability
of both barriers, resulting in an overlap of the wave functions
inside the barrier. Although Slonczweski's model is more realistic,
it does not account for the temperature and voltage dependence of
the TMR ratio. Vacuum tunnel barriers give MR with very little
$V$-dependence \cite{2002-Wulfhekel-JMMM}. Based on this result,
two-step tunnelling through localized states in the tunnel barrier
has been put forward as a possible explanation for the $V$- and
$T$-dependence, as well as for negative TMR values
\cite{2003-Tsymbal-JPCM,1998-Zhang-JAP,2003-Tsymbal-PRL}.

Room-temperature TMR ratio's of several hundreds percent have been
realized \cite{Yuasa,Parkin,Ohno}, sufficiently large to make TMR
hard disk read heads \cite{Seagate} and Magnetic Random Access
Memory (MRAM) \cite{Freescale} commercially attractive. Since TMR
relies on tunnelling through the NM layer, and not on transport as
in GMR (see next section), one can apply insulating organic layers
as spacer. A SAM of alkanethiols has for example been used for this
purpose \cite{Petta}.

\subsection{Giant magnetoresistance}
\label{GMR}

The basic layout of a GMR device was already referred to in the
Introduction. Analogous to the MTJ discussed above, an external
magnetic field is used to switch the relative magnetic
orientations of the FM layers from P to AP, or vice versa. The P
configuration usually, but not necessarily, has a lower
resistance. Although the working of the device seems relatively
simple, the GMR effect was not predicted and its underlying
principles are not straightforward. Before it was explained that
TMR is \emph{directly} related to the DOS asymmetry between the FM
contacts on both sides of the tunnel barrier. Here, we will see
that GMR is also related to a different DOS for both spin
subbands, but in a more indirect fashion. As in the case of TMR,
we assume that spin-flip scattering (i$.$e$.$ the change of
spin-up to spin-down, or vice versa) is negligible:
$\tau_{\uparrow \downarrow}, \tau_{\downarrow \uparrow}
\rightarrow \infty$. This turns out to be a very good
approximation on the timescale of the dissipative processes that
give rise to electrical resistivity \cite{Gregg}. The lack of
interchange between both spin species makes it possible to treat
their transport in terms of two independent transport channels, a
model referred to as the \textit{two-channel model} introduced by
Mott \cite{Mott2channel,Mott2channel2,Mott2channel3}. We make use
of this two-channel model to explain the origin of GMR in the two
existing geometries described as current-in-plane (CIP) and
current perpendicular to plane (CPP) (see figures \ref{CIPGMRfig}
and \ref{CPPGMRfig}, respectively). We assume that all conductors
are in the diffusive limit, i$.$e$.$ the electron mean free path
is much shorter than the typical dimensions of the conductors.
This assumption normally holds for organic conductors. However, in
the case of CNTs transport can be ballistic, see section
\ref{CNT}.

\subsubsection{Current-in-plane GMR}
\label{CIP GMR}

The first GMR devices had the CIP geometry, as they were easier to
fabricate. In a FM metal, the (usually $d$) spin subbands are
split by the exchange interaction (see figure \ref{TMR}),
resulting in a finite magnetization at thermal equilibrium, and in
a different DOS and Fermi velocity for spin-up and spin-down
electrons. As a consequence, both spin species generally have
different bulk conductivities, $\sigma$. In this Review, we define
spins oriented in the direction of the magnetization as the
\emph{majority} carriers, and spins oriented opposite to the
magnetization as the \emph{minority} carriers. The current in a FM
metal is mostly carried by the electrons with the highest
conductivity, normally the majority electrons, and is thus
spin-polarized. The bulk current polarization $\alpha$ of a FM
metal is defined as

\begin{equation}
\alpha =
\frac{\sigma_{\uparrow}-\sigma_{\downarrow}}{\sigma_{\uparrow}+\sigma_{\downarrow}}
\end{equation}

\begin{figure}[htbp]
\center{\includegraphics[width=8cm]{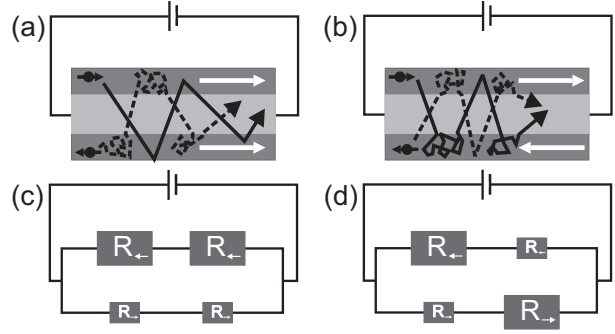}}
  \caption{Schematic representation of a CIP GMR device consisting of two
  FM electrodes (dark grey) and a NM spacer (light grey) for the P (a) and
  AP (b) configuration. The magnetization is denoted by the white arrows.
  The trajectory of two electrons with opposite spin direction is represented
  by the black solid and dotted lines. The corresponding resistor model is
  given for the P (c) and AP configuration (d). A bigger resistor represents
  a larger resistance for the corresponding spin species due to more scattering.}
  \label{CIPGMRfig}
\end{figure}

In a CIP GMR device (see figure \ref{CIPGMRfig}), scattering is
weak for electrons with spin parallel to the magnetization of the
FM layer in which scattering takes place (they are majority
carriers in this layer), whereas scattering is strong for
electrons with opposite spin. Each FM layer in the CIP geometry
thus favours majority carriers. When both FM layers are aligned
parallel [figure \ref{CIPGMRfig}(a)], the resistivity of the spin
channel with spins aligned with the magnetization of both FM
contacts is low (and the resistivity of the other spin channel
high), resulting in an overall low resistance [figure
\ref{CIPGMRfig}(b)]. For antiparallel alignment [figure
\ref{CIPGMRfig}(c)], carriers in both spin channels experience
considerable scattering, resulting in an overall larger resistance
[(figure \ref{CIP GMR}(d)]. The critical length scale in a CIP GMR
device is the \emph{electron mean free path}. For a sizeable
effect, the NM spacer layer should be thinner than the electron
mean path, and the FM layers should be thinner than the mean free
path of the carriers with majority spin.

\subsubsection{Current-perpendicular-to-plane GMR}
\label{CPP GMR}

\begin{figure}[htbp]
\center{  \includegraphics[width=4in]{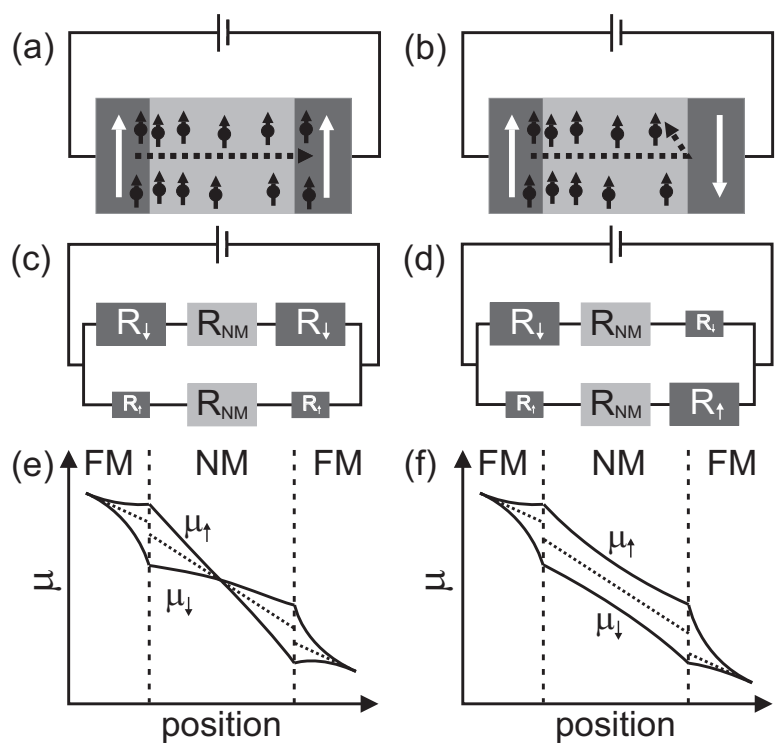}}
  \caption{Schematic representation of a CPP GMR device consisting of
  two FM electrodes (dark grey) separated by a spacer (light grey)
  for the P (a) and AP configuration (b). The magnetization of the
  FM electrodes is denoted by the white arrows. The dotted arrows
  represent the spin current. The corresponding resistor model is
  given for the P (c) and AP configuration (d). The colours correspond
  to the layers in (a) and (b), and bigger resistors represent a
  larger resistance for the denoted spin species. The electrochemical
  potentials $\mu$ for the two spin species are given for the P (e)
  and AP configuration (f). The dotted lines are the asymptotes of
  the electrochemical potentials to which they would collapse at
  large distances. The dashed lines correspond to the interfaces in
  (a) and (b).}
  \label{CPPGMRfig}
\end{figure}

The CPP GMR geometry of figure \ref{CPPGMRfig} is most commonly used
in organic spintronic devices. The physical origin of CPP GMR is
rather different from that of CIP GMR.
When a FM contact is connected to a NM material and a current is
driven through the system, the spin-up current is different from the
spin-down current, due to the current polarization in the FM. A
finite magnetization builds up in the NM material, which is known as
\textit{spin accumulation} \cite{JohnsonSilsbee}. The spin
accumulation is defined as the difference between the
electrochemical potential for spin-up electrons, $\mu_{\uparrow}$,
and that for spin-down electrons, $\mu_{\downarrow}$. The magnitude
of the spin accumulation depends on the spin injection rate into the
normal material and the spin relaxation time, and it decays
exponentially away from the injecting contact on a length scale set
by the spin relaxation length
\begin{equation}
\mu_{\uparrow} - \mu_{\downarrow} \propto \exp(-l/l_s),
\end{equation}
where $l$ is the distance from the injecting contact. The net spin
density resulting from the spin accumulation is typically orders of
magnitude smaller than the charge density in the NM. However, the
spin accumulation in the NM can be probed by a second FM contact,
the spin detector, if it is placed at a distance smaller or
comparable to the spin relaxation length from the spin injector.

A finite spin accumulation implies different densities of spin-up
and spin-down carriers at the site of the detector interface. The
transmission is now largest when the magnetization of the detector
contact is parallel to the net spin accumulated at its interface.
CPP GMR can also be described in terms of a parallel resistor model,
as shown in figures \ref{CPPGMRfig}(c) and (d). A more thorough
theoretical description of CPP GMR based on the Boltzmann equation,
has been provided by Valet and Fert, for which we refer to
\cite{ValetFertmodel}. With their model, the electrochemical
potentials of the two spin species can be calculated, as illustrated
in figure \ref{CPPGMRfig}(e) and (f). It reveals the splitting of
the electrochemical potentials at the interfaces of the FM contacts
and non-magnetic material. It also shows the different voltage drop
(represented by the discontinuity of the asymptote) at the
interfaces for the P and AP configuration, which leads to the
difference in resistance between these two cases. It is important to
note that the critical length scale for CPP GMR devices is the
\emph{spin relaxation length}, and not the electron mean free path
as for CIP GMR. As the spin accumulation decays exponentially from
the injector, the GMR ratio depends exponentially on the distance
between injector and detector. This feature is very useful for
determining the spin relaxation length in (organic) materials.

Besides the basic trilayer CIP and CPP geometries described above,
GMR has been observed in multilayer systems \cite{Baibich}, granular
systems \cite{Black} and nanocontacts \cite{Jedema}.

\subsection{Conductivity mismatch problem}
\label{mismatch}

A fundamental obstacle for spin injection from a FM metal into a
semiconductor is the so-called \textit{conductivity mismatch
problem} \cite{Fert2,Schmidt0,Schmidt1,Rashba,Smith,Yu}. The
conductivity of a semiconductor is usually much lower than that of
a metal. In a SV, one likes to detect the resistance change due to
the different magnetization orientations in the FM layers. If the
resistance of the whole device is dominated by the resistance of
the semiconductor spacer, the overall resistance change is
negligible. This can also be seen from the resistor model in
figure \ref{CPPGMRfig}. When the resistance of the NM material,
$R_{NM}$, is much larger than the other resistances, this
dominates the overall resistance and no change is observed. This
is particularly relevant for organic spintronics, since most
organic materials are much less conductive than the FM contacts.

There are two possible solutions to this problem. The first one is
to use a fully spin-polarized FM material, i$.$e$.$ a
\emph{half-metal} such as LaSrMnO$_3$ (LSMO) \cite{Park,Bowen}. In
the classical ferromagnetic materials (e$.$g$.$ Fe, Ni, Co), the
conduction electrons mainly have 4s character, whereas the polarized
electrons are in the, more localized, 3d band. This electronic
structure leads to a spin polarization at the fermi level far below
100\%: Co is the best elemental ferromagnet with $P$=45\%
\cite{1994-Meservey-PhysRep}. In a half-metal, only one spin subband
is occupied at the Fermi level and the spin polarization $P$
therefore approaches 100\% at low temperatures. In the case of LSMO
($T_C \sim 370$ K), there is a fully polarized conduction band of 3d
character at $E_F$, and no s band. Even if the bulk properties of a
material indicate half-metallic behaviour, it is not \emph{a priori}
clear however whether the spin polarization can efficiently be
transferred across the interface with a NM material. The maximum
contact spin polarization value observed in MTJs with LSMO is 0.95
\cite{Bowen}. LSMO contacts have been applied in spin valve devices
with CNTs and organic thin films (see sections \ref{CNT exp} and
\ref{OTF exp}, respectively).

\begin{figure}[htbp]
\center{  \includegraphics[width=8cm]{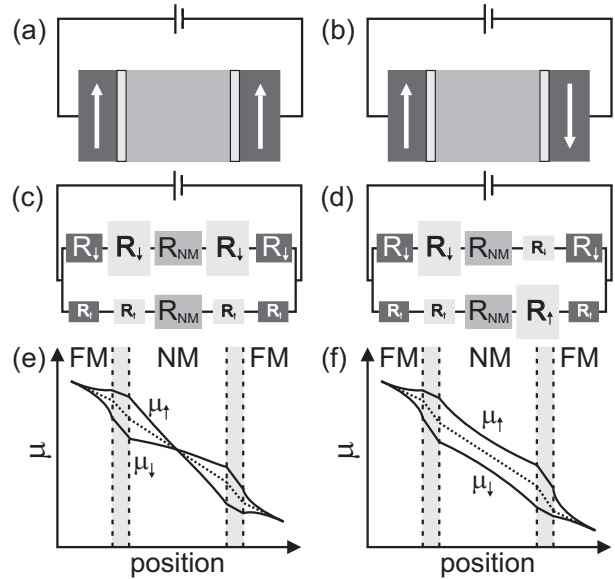}}
  \caption{Schematic representation of a CPP GMR device consisting
  of two FM electrodes (dark grey) separated by a spacer (light grey)
  and tunnel barriers (light grey with black outline) for the P (a)
  and AP configuration (b). The corresponding resistor model is given
  for the P (c) and AP configuration (d). Colours correspond to the
  different parts in (a) and (b). The electrochemical potentials
  $\mu$ for the different spin species are given for the P (e) and
  AP configuration (f). The dotted lines are the asymptotes of the
  electrochemical potentials to which they would collapse at large
  distances. The dashed lines correspond to the interfaces in (a)
  and (b).}
  \label{mismatchfig}
\end{figure}

Another possible solution for the conductivity mismatch problem is
the introduction of a large spin-dependent resistance
\cite{Fert2,Rashba,Smith,Yu}. This spin-dependent resistance could
be a tunnel barrier in between the FM contact and semiconductor
spacer. This spin-dependent resistance gives a larger change in
resistance between the P and AP configuration, as can be
visualized by the resistor models in figure \ref{mismatchfig}(c)
and (d). In the model by Valet and Fert this will lead to a larger
spin splitting at the interface and a bigger difference in the
voltage drop over the whole device for the two configurations. See
also figure \ref{mismatchfig}(e) and (f).

In the case of organic spacers, next to the bulk conductivities of
the FM contacts and the organic material and the interface
resistances, also the ratio of polarons to bipolarons is of
importance. Bipolarons have no spin and the spin-polarized current
is only carried by polarons. Ren \textit{et al.} \cite{Ren} find,
like in the case of inorganic semiconductors, an increase of the
spin polarization when the conductivity of the organic material
approaches or surpasses the conductivity of the FM material and when
a spin-dependent interface resistance is introduced. The influence
of the bipolarons is not drastic. When there are only bipolarons the
spin polarization is zero of course, but when the fraction of
polarons is only 20 \%, the spin polarization is 90 \% of the value
attainable with only polarons and no bipolarons.

\subsection{Spurious effects}
\label{spurious}

Injecting and detecting spins in a NM material is not trivial, as is
apparent from the discussion of the conductivity mismatch problem
above. In this section, we discuss a number of phenomena (or
`spurious effects') that can give rise to MR effects, but are not
related to (but are easily mistaken for) the TMR and GMR effects
described above. For the correct interpretation of organic
spintronic experiments, it is crucial to take these effects into
account.

The Lorentz force curves the electron trajectories and has a
(positive) MR effect on the order of $(l_e/l_B)^4$, where $l_B =
\sqrt{h/(eB)}$ is the magnetic length \cite{2006-prigodin-synmet}.
Lorentz magnetoresistance (LMR) is relevant for systems with a
relatively large mean free path, such that $\omega_c \tau > 1$,
where $\omega_c$ is the cyclotron frequency and $\tau$ the elastic
scattering time \cite{Yang99}. In systems where transport takes
place by hopping, a magnetic field can enhance the localization of
the carriers on the hopping sites, thereby also increasing the
resistance \cite{2005-wohlgenannt-PRB}.

In the coherent, diffusive transport regime, conductance can be
affected by a magnetic field via electron interference phenomena
such as weak localisation (WL) and universal conductance
fluctuations (UCF) \cite{BeenakkerVanHouten}. WL is interpreted as
coherent backscattering and gives rise to an enhanced resistance
around $B = 0$, where the width of the resistance maximum is
determined by the (charge) coherence length, $l_{\phi}$. UCF are of
the order $e^2/h$, regardless of the sample size and the degree of
disorder (hence the name `universal'). UCF result from the
microscopic change in electron interference paths due to a change in
$E_F$, impurity configuration or enclosed magnetic flux.

Another MR effect that is not caused by spin accumulation in the NM
material is the \emph{local Hall effect}. Stray fields coming from
the FM contacts can penetrate the NM spacer and induce local Hall
voltages. When the magnetization of the FM contacts changes, so do
the Hall voltages \cite{Monzon1,Monzon2}. In this way, these
voltages can obscure the true spin valve signal.

In small systems, such as CNTs, where Coulomb charging effects are
relevant, the \emph{magneto Coulomb effect} (MCE) can play a role
\cite{vanderMolen}. Due to this effect, the conductance in a system
connected to two FM leads changes as a function of magnetic field,
but this is not caused by spin accumulation. In a FM contact, the
spin subbands are shifted by the Zeeman energy in opposite direction
under the influence of an external magnetic field. As the DOS at
$E_F$ in a ferromagnet is in general different for both spin
species, repopulation of the electrons takes place through spin-flip
scattering. This gives a shift in the chemical potential
\cite{Shimada}. When the FM contacts are connected to big NM leads,
the change in chemical potential in the FM contacts causes electrons
to flow across the FM/NM interface. This leads to a change in the
dipole layer at the interface. The voltage change can couple to the
conductor in between the two FM leads via the capacitance and
therefore effectively acts like a gate. As the magnetization of the
FM material switches its direction at the coercive field, the
conductance changes discontinuously at this field due to the MCE.
The MCE can therefore easily be mistaken for the SV effect in small
structures \cite{vanderMolen}.


\begin{figure}
\center{  \includegraphics[width=6cm]{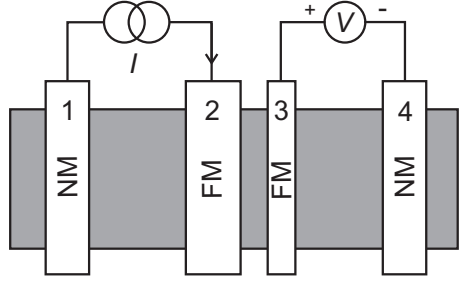}}
  \caption{Non-local geometry for measuring spin accumulation.
  A current is driven between electrodes 1 and 2, while the voltage
  is measured between electrodes 3 and 4. When electrode 2 is FM, a
  spin-polarized current is injected. The spins will diffuse in both
  directions and can therefore be probed by electrode
  3, which is also FM. Therefore a voltage difference can be observed
  between the parallel and antiparallel state of the two FM electrodes.
  Because the measured polarization is not influenced by the current,
  the signal is purely due to spin transport.}
  \label{nonlocal}
\end{figure}

In principle, it is possible to calculate the magnitude of the above
spurious effects or measure them in control devices \cite{Bird}.
However, a more elegant and rigorous way to rule out the discussed
effects is to measure spin accumulation using the so-called
\emph{non-local geometry} of figure \ref{nonlocal} \cite{Jedema}. A
current is injected by two contacts, while the voltage is measured
by two other contacts. When at least two contacts are FM, the spin
accumulation can be probed in such a way that the measured spin
diffusion is isolated from the current path. See for an example of
such a measurement figure \ref{nonlocal}. In this way, the measured
signal is only due to spin accumulation.

\section{Spin relaxation}
\label{spinrelax}

In general, one can distinguish two spin relaxation processes. The
first one describes the decay of a net spin component along the
axis of spin quantization, which we define as the $z$-axis. The
$z$-component (or \emph{longitudinal} component) of the total
spin, $S_z$, decays exponentially to equilibrium due to individual
spin flips on a timescale $T_1$. This $T_1$ is equal to the spin
relaxation time $\tau_s$, defined in (\ref{reltime}). As this
process requires energy exchange with the environment, it is a
rather slow process. There is a second process, however, that does
not require energy exchange and affects the spin component
perpendicular to the quantization axis, i$.$e$.$ the
\emph{transverse} component $S_{\perp}$. This process affects the
quantum-mechanical phase of individual spins and leads to loss of
\emph{coherence} on a timescale $T_2$. For different spins within
an ensemble the phases are in general affected unequally, which
results in the spins getting out of phase. The timescale related
to this process of ensemble dephasing is often denoted as $T_2^*$
\cite{Kikkawa2000,Khaetskii2002,Golovach2004}. Usually $T_2^* <
T_2$, an effect referred to as \emph{inhomogenous broadening}. The
time evolution of a spin ensemble with total spin
\textbf{S}(\footnote{We use the symbol \textbf{S} for the
resultant spin vector of a spin ensemble, whereas
$\overrightarrow{S}$ is used for denoting an individual spin
vector.}) in an external magnetic field \textbf{B} along the
$z$-axis can then be described by the Bloch equations
\begin{eqnarray}
  \frac{dS_z}{dt} &=& \gamma (\mathbf{B} \times \mathbf{S})_z - (S -
S_z)/T_1 \\
 \frac{dS_{\perp}}{dt} &=& \gamma (\mathbf{B} \times
\mathbf{S})_{\perp} - (S - S_{\perp})/T_2^*,
\end{eqnarray}
where $\gamma$ is the gyromagnetic ratio.

In this section we discuss the underlying mechanisms for spin
relaxation in solids, divided in mechanisms related to spin orbit
coupling and to hyperfine interaction. Both spin orbit coupling and
hyperfine interaction are expected to be small, but not completely
negligible for most organic materials. The dominant relaxation
mechanisms in organic materials are still rather unclear. There are
a few reports where the spin relaxation length is determined from
fitting to Julli\`{e}re's formula, but it is hard to distinguish
between spin relaxation at the interfaces and within the organic
material itself. Also, the simple Julli\`{e}re formula
(\ref{Julliere}) is not always very appropriate for the applied
device configurations.

\subsection{Spin-orbit coupling}
\label{spin orbit}

Spin-orbit coupling is a relativistic effect, describing the
interaction between the electron's spin and its orbital motion
around an atomic nucleus. More generally, spin-orbit coupling occurs
whenever a particle with non-zero spin moves in a region with a
finite electric field. In the rest frame of a particle moving at a
relativistic velocity, a static electric field Lorentz-transforms in
a field with a finite magnetic component. Thus, although the spin
degree of freedom only couples to a magnetic field, it is indirectly
affected by an electric field via spin-orbit coupling. The
electrical field can have various physical origins, such as the
electric field of an atomic nucleus or the band structure of a solid
\cite{Filip}.

As spin-orbit coupling generally grows with atomic number $Z$ (it
scales as $Z^4$ in the case of an hydrogen-like atom
\cite{Sanvito}), and organic materials consist mainly of low-$Z$
materials (in particular C), spin-orbit coupling is usually small in
organic materials. Sulphur atoms could provide a considerable
spin-orbit coupling, but these atoms normally play a marginal role
in carrier transport in organic materials
\cite{1996-Bennati-JChemPhys}. Depending on the exact band structure
of the organic material, spin-orbit coupling is actually not always
negligible \cite{Reinder}.

In (inorganic) solids one can distinguish two main contributions to
spin-orbit coupling. The first one, termed the \emph{Dresselhaus
contribution}, occurs in crystals that exhibit bulk inversion
asymmetry, which implies that there is a net electric field for
certain crystal directions \cite{Dresselhaus,Perel}. The second one,
referred to as the \emph{Rashba contribution}, occurs in systems
with net electric fields due to structural inversion asymmetry
\cite{Rashba,Bychkov}. Three different spin-orbit-coupling-related
spin relaxation mechanisms can be distinguished in non-magnetic
solids: Elliot-Yafet (EY), D'yakonov-Perel (DP), and
Bir-Aronov-Pikus (BAP).

The EY mechanism \cite{Elliot} is due to the fact that under the
influence of spin-orbit coupling momentum eigenstates are no spin
eigenstates anymore. Any momentum scattering event has hence a
finite probability to flip the spin. The EY mechanism leads to a
spin relaxation time $\tau_s$ that is proportional to the momentum
scattering time. Momentum scattering is mainly caused by impurities
at low temperature and phonons at high temperature \cite{Zutic}.
Usually EY is the dominant mechanism in metals, and it has been
recently claimed to be dominant in organic semiconductors as well
\cite{Pramanik}.

The DP \cite{Perel} mechanism arises when the solid lacks a center
of symmetry, and is therefore directly related to the Dresselhaus
contribution. As the internal magnetic field is
$\overrightarrow{k}$-dependent, the axis around which the spin
precesses is randomized upon electron (momentum) scattering. This
results in a loss of memory of the initial spin direction. Heavy
scattering slows down the spin relaxation, because the spin cannot
follow the internal magnetic field when it changes too rapidly.
Therefore, the spin relaxation time is inversely proportional to the
scattering time.

The BAP \cite{Bir} mechanism is caused by electron-hole exchange
interaction, and therefore only plays a role in systems where there
is a large overlap between the electron and hole wave functions.

The spin-orbit-coupling-related relaxation mechanisms directly
affect $T_1$, and indirectly $T_2$.

\subsection{Hyperfine interaction}
\label{hyperfine}

Another source for spin relaxation is the hyperfine interaction. It
originates from the interaction of the electron spin with the
nuclear spins of the host material. In general, the electron spin
interacts with many, say $N$, nuclear spins. The electron-nuclear
coupling Hamiltonian is then given by
\begin{equation}
H_{hyp} = \sum_i^N A_i \overrightarrow{I_i}\cdot \overrightarrow{S},
\end{equation}
where $\overrightarrow{I_i}$ and $\overrightarrow{S}$ are the spin
operator for nucleus $i$ and the electron spin, respectively, and
$A_{i}$ the coupling strength between them.

The nuclear spins affect the spin relaxation time $T_1$ by means of
so-called electron-nuclear flip-flops. In addition, fluctuating
nuclear spins also results in dephasing, thus affecting $T_2$. For
an electron spin interacting with $N$ nuclear spins, the statistical
fluctuation scales with $\frac{1}{\sqrt{N}}$
\cite{Merkulov,Khaetskii}. Hence the more delocalized the electron
wave function is, the less influence of the nuclei.

The nuclear spins in organic materials are mainly originating from
the isotopes $^1$H ($I$ = 1/2), $^{13}$C ($I$ = 1/2), and $^{14}$N
($I$ = 2). Despite the presence of nuclear spins, the hyperfine
interaction in organic materials is usually weak. The reason is that
for organic conductors often use is made of $\pi$-conjugated
molecules with delocalized states (see section \ref{orgelec}) that
have practically no overlap with the C or H atoms \cite{Sanvito}.



\section*{PART II}

\section{Carbon nanotube devices}
\label{CNT exp}

In this section we discuss experiments on organic spin valves where
the spacer between the FM electrodes is formed by a CNT, including
tunnelling or diffusive junctions. Generally, the CNT is considered
to be a quasi-spin-ballistic waveguide. Where purely metallic
systems (such as the GMR multilayer systems discussed in section
\ref{CPP GMR}) have the advantage of large carrier velocity,
$\tau_s$ is very short ($\sim$10$^{-10}$ s). (Organic)
semiconductors, on the contrary, have much larger $\tau_s$ (up to
$\sim$10$^{-6}$ s), but the carrier velocities are smaller. CNTs
combine a high carrier velocity ($\sim$10$^6$ m/s) with a
potentially very long $\tau_s$. It is therefore no surprise that the
field of organic spintronics took off with work on CNTs, and that
there is still a lot of activity going on.

\subsection{Multi-wall carbon nanotubes}
\label{MWCNT exp}

\begin{figure}[htbp]
\center{ \includegraphics[width=7cm]{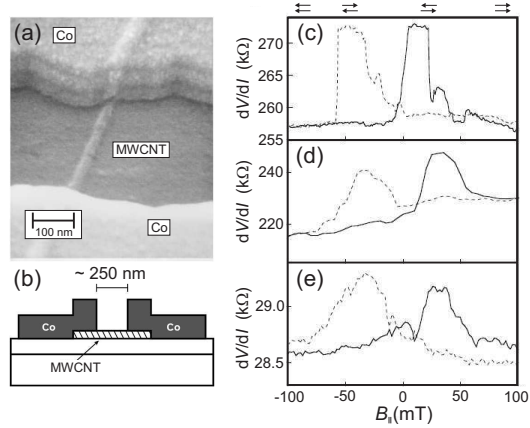}}
  \caption{(a) Scanning electron microscope picture and (b) schematic
  diagram of a device, consisting of a MWCNT connected by two 65 nm
  thick polycrystalline Co contacts, fabricated by EB
  lithography and thermal evaporation. The contacts lie on top of the
  MWCNT and are separated by 250 nm. The diameter of the MWCNT in
  this particular device is 30 nm (diameters range from 10 - 40 nm)
  and the length more than 1 $\mu$m. Non-FM leads are deposited more
  than 30 $\mu$m away from the MWCNT. The device is fabricated on a
  semi-insulating Si wafer covered by 200 nm SiO$_2$. The two-terminal
  differential resistances of three different MWCNT devices at a
  temperature of 4.2 K are given in (c), (d) and (e). The magnetic
  field is pointing parallel to the substrate and the obtained values
  for the MR are (c) 6\%, (d)9\% and (e)2\%. The arrows at the top of
  the graph denotes the magnetization of the left and right contact. Courtesy of K. Tsukagoshi\cite{1999-Tsukagoshi-Nature}.}
  \label{tsukagoshi}
\end{figure}

The first organic spintronic device ever was realized by Tsukagoshi
and conworkers \cite{1999-Tsukagoshi-Nature}. It consisted of a
single MWCNT contacted by polycrystalline Co contacts, see
figure\ref{tsukagoshi}(a). The device layout is schematically given
in figure \ref{tsukagoshi}(b). The MWCNTs -- 10-40 nm in diameter
and about 1 $\mu$m long -- were synthesized by the arc discharge
evaporation method in a He atmosphere to avoid contamination by
magnetic impurities (e$.$g$.$ from catalyst particles). The Co top
contacts are defined by electron-beam (EB) lithography and thermal
evaporation.

Examples of MR measurements at 4.2 K are given in figures
\ref{tsukagoshi}(c)-(e). There is a rather large sample-to-sample
variation in the differential resistance, most probably due to
irreproducible contact resistances. CoO could for example be formed
at the interface. Residues from the resist layer cannot be excluded
either. A hysteretic resistance increase of $\sim$50 mT width is
observed around $B = 0$, implying spin-valve behaviour. The two Co
contacts are nominally the same and should therefore have the same
coercive field. The authors nevertheless argue that AP alignment is
possible due to local magnetization fluctuations on the scale of the
MWCNT diameter (30 nm). A maximum 9\% MR ($MR \equiv (R_{AP} -
R_P)/R_{P} \approx (R_{AP} - R_P)/R_{AP}$) is reported for these
devices at $T$ = 4.2 K. The MR decreases approximately exponentially
with $T$ and disappears around 20 K. The $T$-dependence is ascribed
to the poor quality of the MWCNT/FM interface. In later reports on
the same device structure
\cite{2000-Tsukagoshi-Superlattices,2003-Chakraborty-APL}, it was
observed that the MR becomes negative above 20K before completely
disappearing at 175 K. The negative MR is possibly due to the
presence of CoO \cite{2003-Chakraborty-APL} or due to the negative
spin polarization of the Co d-band, where the majority spin DOS at
the Fermi level is smaller than the minority spin DOS.

Comparing with a simple approximation based on Julli\`{e}re model
and neglecting the influence of the MWCNT-Co interfaces, Tsukagoshi
\textit{et al.} find $l_s \sim$ 130 nm
\cite{1999-Tsukagoshi-Nature}. As this estimate is based on only one
distance between the FM contacts (250 nm), it does not take into
account spin relaxation at the interfaces. The estimate is therefore
likely to be a conservative one. Stray magnetic fields from the FM
contacts can affect the resistance of the MWCNT by suppressing weak
localisation (see section \ref{spurious}). For this experiment the
difference in stray fields between the AP and P configuration is
estimated to be responsible for a maximum MR of 0.3\%
\cite{2001-Alphenaar-JAP}, hence smaller than the observed maximum
of 9\% in \cite{1999-Tsukagoshi-Nature}. For comparison, devices of
the same geometry but with one FM (Co) contact and one NM contact
(Pt/Au) were also measured, showing no hysteretic MR contrary to the
former devices \cite{2001-Alphenaar-JAP}.

Similar devices have been tested again by this group
\cite{2003-Chakraborty-APL} and other groups
\cite{2001_Orgassa_nanotech, 2002-schneider-APL, 2002-schneider-JAP,
2005-ishibashi-PSS}. Although qualitatively the same MR behaviour is
repeatedly reported, there is a major lack of consistency in the
reported quantities.


\begin{figure}[htbp]
\center{ \includegraphics[width=5cm]{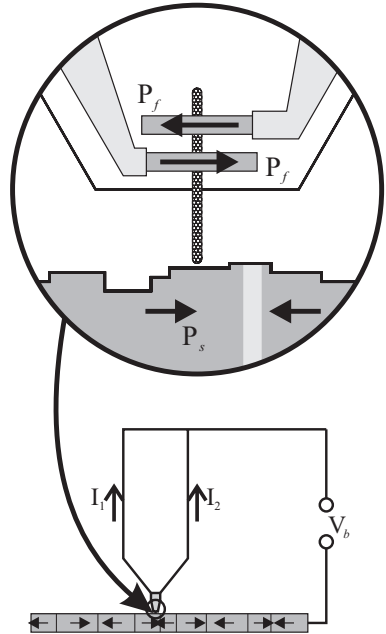}}
  \caption{Setup for spin-resolved scanning tunnelling microscopy
  with a CNT tip, as proposed by Orgassa \textit{et al}$.$ The spin
  polarization in the contacts P$_f$ determines the asymmetry in the
  currents I$_{1,2}$. Courtesy of D. Orgassa\cite{2001_Orgassa_nanotech}.}
  \label{orgassa}
\end{figure}

An alternative contacting method, involving shadow mask
evaporation was introduced by Orgassa \textit{et al.}
\cite{2001_Orgassa_nanotech}. They have used a 4 $\mu$m tungsten
wire as a shadow mask over an orthogonally placed nanotube and
evaporated Co and NiFe electrodes from two different angles,
resulting in a 1 $\mu$m contact spacing. A maximum 2.2\% negative
and 0.6\% positive MR was reported in two out of ten Co/MWCNT/NiFe
devices below 30K. Comparing the MR results with
Julli$\textrm{\`{e}}$re model and using spin polarizations in the
contacts taken from literature, gives 380 nm as the lower limit
for spin scattering length in MWCNT. Similar to Tsukagoshi
\textit{et al}$.$, this estimate is only based on one contact
spacing and therefore does not account for spin relaxation due to
interface imperfections. It is likely that also in this experiment
the growth conditions of the electrodes dominate the device
performance. Orgassa \textit{et al}$.$ have also proposed a
concept for spin-resolved scanning tunnelling microscopy (SR-STM),
based on a CNT acting as the tunnelling tip in proximity to a
magnetic sample \cite{2001_Orgassa_nanotech}.



A very large increase in MR was realized by Zhao \textit{et al.},
reporting 30\% \cite{2002-schneider-APL} and -36\%
\cite{2002-schneider-JAP} MR in Co/MWCNT/Co devices at small bias
currents at 4.2 K. Their fabrication method is very much like
Tsukagoshi's and a wide range of room temperature resistance of the
devices is reported here as well. The MR signal decreases with bias
current and disappears above 10K. At low temperature non-linear
transport is observed, possible caused by Coulomb blockade in the
$\sim$200 nm long devices.


Considerable improvement in realizing reliable, low-ohmic contact
resistance was achieved by the introduction of
$\mathrm{Pd_{0.3}Ni_{0.7}}$ electrodes by Sahoo \textit{et al.}
\cite{2005-sahoo-APL}. The contact resistance can be as low as 5.6
K$\Omega$ at 300 K. While the Pd alloy is expected to have the same
contact properties -- such as low-resistance and quasi-adiabatic
contacts -- as pure Pd contacts to CNTs \cite{2003-guo-nature}, the
high Ni concentration provides the required spin current. The spin
polarization in $\mathrm{Pd_{0.3}Ni_{0.7}}$ is estimated 9.58 \%,
and the $\mathrm{Pd_{0.3}Ni_{0.7}}$ thin-film Curie temperature and
saturation magnetization are half the bulk values
\cite{2005-sahoo-APL}. This is ascribed to partial oxidation of the
Ni during evaporation. A low device resistance, and in particular a
low contact resistance, is not required a priori in a spintronic
device, as was pointed out in section \ref{mismatch}. However, low
contact resistance avoids charging effects and hence the
magneto-Coulomb effect, see section \ref{spurious}. A maximum MR
value of $\sim$2\% is reported at 1.8 K and 2 V back gate voltage.
The MR is however strongly dependent on the gate voltage and
disappears at zero gate voltage, which was not understood.


\begin{figure}[htbp]
\center{\includegraphics[width=8cm]{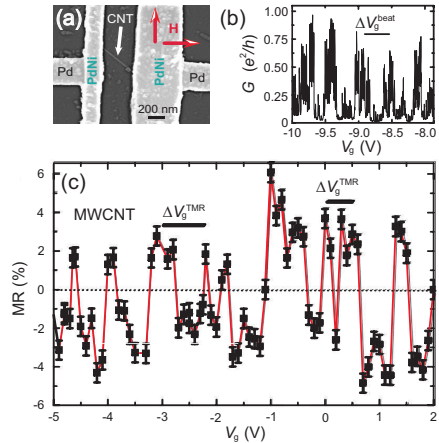}}
  \caption{(a) SEM image of a CNT connected to ferromagnetic PdNi contacts.
  The separation between the contacts is 400 nm. The magnetic field is
  applied parallel to the substrate, denoted by the arrows on the top right.
  (b) The linear conductance as a function of the gate voltage, measured at
  300 mK. The bars reflect the error of the measurement. This figure shows
  a beating pattern with $\Delta V_g^{beat} \approx 0.4$V. (c) MR values
  as a function of gate voltage for a MWCNT, measured at 1.85K. The
  oscillations have a typical scale of 0.4 to 0.75 V, roughly corresponding
  to the beating pattern in (b). Courtesy of S. Sahoo\cite{2005-sahoo-natphys}.}
  \label{SahooMWCNT}
\end{figure}

The gate voltage dependence of the MR was worked out in more detail
in an experiment by the same group \cite{2005-sahoo-natphys}, where
the MR magnitude and sign could be gate-field-tuned in a predictable
way, see figure \ref{SahooMWCNT}(a)-(c). The MR varies here almost
regularly between -5\% and +6\% on a gate voltage scale of $\sim$0.5
V at 1.85 K. The MR oscillation cannot be explained in terms of
Rashba spin-orbit coupling \cite{Bychkov}, as proposed by Datta and
Das \cite{Datta}, since the spin-orbit coupling is too low in CNTs
(see section \ref{spin orbit}). Alternatively, the oscillatory
behaviour was shown to be consistent with quantum interference as
predicted originally for semiconductor heterostructures
\cite{Schapers}. As was shown in measurements at lower temperature
($T$ = 300 mK), the MWCNT behaves as a quantum dot
\cite{LeoNATO,Wiel}. Weak disorder in the MWCNT causes the
single-electron resonances to be modulated in amplitude, see figure
\ref{SahooMWCNT}(b). The MR signal [figure \ref{SahooMWCNT}(c)] is
claimed to follow this envelope function, which is substantiated in
the discussion of their SWCNT experiment in section \ref{SWCNT exp}.


In the experiments discussed so far, metallic FM contacts were
attached to the MWCNTs. Hueso \textit{et al.} \cite{2006-Hueso-APL}
instead applied single-crystal La$_{2/3}$Sr$_{1/3}$MnO$_3$ (LSMO),
which is believed to be half-metallic at low temperatures and to
remain ferromagnetic at room temperature (see section
\ref{mismatch}). The 30 nm thick LSMO contacts are fabricated by
pulsed laser deposition (PLD) and focused ion beam (FIB) milling.
MWCNTs are then dispersed on the patterned substrate from solution
and successfully contacted devices are selected. The
room-temperature contact resistance of LSMO-CNT-LSMO devices is
compared to Pd-CNT-Pd control devices, and is found to be twice as
high. The LSMO-CNT-LSMO devices show a conductance gap around zero
bias voltage below 200K, saturating to 250 mV at low temperature.

Very recently, Hueso \textit{et al}$.$ \cite{2007-hueso-Nature}
reported a maximum MR ratio of 61 \% in a single MWCNT situated on
top of two LSMO electrodes at 5 K. They find a spin lifetime of 30
ns and a spin diffusion length of 50 $\mu$m. The MWCNT-LSMO
interfaces behave like tunnel barriers, which is favourable for the
spin signal (see section \ref{mismatch}). The tunnel barriers also
limit the current, which allows for large-bias ($\sim$25 mV)
measurements. The relatively large bias voltage circumvents the
occurrence of Coulomb blockade and level quantization effects, and
is a necessary condition for achieving large output signals. Their
MR value corresponds to 65 mV, suitable for applications. The MR
value drops to zero at 120 K, which is at a higher temperature than
in earlier MWCNT devices.

Hoffer \textit{et al.} \cite{2004-hoffer-EPL} have measured MWCNTs
obtained by chemical vapour deposition (CVD) in porous alumina
mebranes. This fabrication method lacks the potential for gating the
devices.
Schneider \textit{et al.} report of a method for fabricating CNTs
filled with FM materials (Co, Fe, Ni) \cite{2004-schneider-DRM}. It
is a magnetic nanowire growth method, as well as a new method for
contacting electrodes to carbon nanotubes in spintronic devices of
this kind.
An alignment method for MWCNTs is proposed by Niyogi \textit{et al.}
\cite{2004-Niyogi-JPCB} in which carbon nanotubes end-capped with
ferromagnetic material can self assemble on predefined ferromagnetic
contacts after introducing magnetic fields. This assembling
technique could be relevant for actually realizing CNT-based
spintronic devices.


\subsection{Single wall carbon nanotubes}
\label{SWCNT exp}

SWCNTs possess some interesting characteristics as compared to
MWCNTs for spin transport studies: less scattering (ballistic
nature), well-defined band structure, and enhanced Coulomb
interaction \cite{CNTbook}. SWCNTs form ideal 1D electronic systems
for studying TLL behaviour (see section \ref{CNT}). On the other
hand, SWCNTs are more difficult to reliably contact than MWCNTs,
because of their smaller diameter and smaller mechanical stability
\cite{1999-Schonenberger-APA,2002-kim-PRB}. SWCNTs are also more
difficult to synthesise in large quantities, which makes them much
more expensive as compared to MWCNTs \cite{Baughman}.

Although \emph{bundles} of SWCNTs (less than 10 nm in bundle
diameter) contacted with Co electrodes have been very briefly
discussed in \cite{2000-Tsukagoshi-Superlattices}, the
interpretation of those data was difficult.
The first report on single SWCNTs was published by Kim \textit{et
al.}, who find a maximum MR of 3.4\% at 0.2 K in Co/SWCNT/Co
\cite{2002-kim-PRB}. The SWCNTs are grown by chemical vapor
deposition \cite{1998-Kong-Nature} on SiO$_2$. Co contacts are
thermally evaporated on top of selected SWCNTs, after which a
rapid thermal annealing is performed. The SWCNT is weakly coupled
to the Co contacts and the authors expect there is a tunnelling
barrier at the Co/SWCNT interfaces. The tunnel coupling results in
quantum dot behaviour, but the size of the Coulomb gap suggests
that the dot size is not determined by the distance between the
contacts but by disorder within the SWCNT. It is mentioned that
SWCNTs with low-ohmic contacts do not show any MR. This could be
due to the conductivity mismatch problem discussed in section
\ref{mismatch}, and thus indicates that the overall transport
through the Co/SWCNT/Co system is diffusive. A spin relaxation
length of 1.4 $\mu$m is estimated at 4.2 K, but also this estimate
is based on only one contact separation (1.5 $\mu$m) and therefore
not very reliable. The MR does not decay with the contact
separation, as the MR values for a 420 nm spacing are smaller than
that of the 1.5 $\mu$m spacing.


Jensen \textit{et al}$.$ fabricated SWCNTs by CVD on catalyst sites
of Fe$_2$O$_3$/Mo supported by Al$_2$O$_3$ nanoparticles placed on a
highly doped SiO$_2$/Si substrate
\cite{2003-Jensen-worldscientific}. The SWCNTs are contacted by Fe
contacts capped with Au. The contact resistance in these devices is
large as compared to devices with pure Au electrodes. They report
100\% MR at 300 mK (60\% at 4.2 K) in a Fe/SWCNT/Fe device
\cite{2003-Jensen-worldscientific}.


\begin{figure}
\center{\includegraphics[width=8cm]{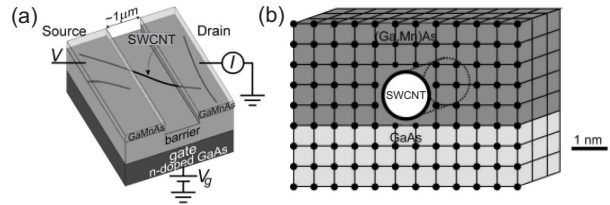}}
  \caption{(a) Schematic picture of a SWCNT connected by two FM
  semiconductor contacts. The SWCNT is lying on a superlattice barrier
  and a heavily n-doped GaAs layer serves as a gate. The SWCNT is connected to two 30-50 nm thick island of Ga$_{0.95}$Mn$_{0.05}$As
  covered with a 3 nm GaAs layer to prevent oxidation. The separation between the two contacts is $\sim$1 $\mu$m. (b) Schematic image of the SWCNT
  encapsulated in the semiconductor crystal. Courtesy of A. Jensen\cite{2004-jensen-nanolett}.}
  \label{Jensen}
\end{figure}

Following their earlier experiments on SWCNTs, Jensen and coworkers
reported on a new fabrication method in which SWCNTs are fully
encapsulated in epitaxially grown heterostructures of GaAs/AlAs and
(Ga,Mn)As, see figure \ref{Jensen} \cite{2004-jensen-nanolett}. This
is the first reported inorganic hybrid semiconductor-CNT structure.
The fabrication starts with an n-doped GaAs substrate (backgate)
followed by a GaAs(2 nm)/AlAs (2 nm) superlattice (gate insulator),
capped with 20 nm of epitaxial GaAs and a layer of amorphous As.
Laser ablated SWCNTs are then dispersed on top of the amorphous As
layer, after which the amorphous As is evaporated in the MBE
chamber, leaving the SWCNTs on a clean GaAs crystal surface. The
SWCNTs are then overgrown by epitaxial Ga$_{0.95}$Mn$_{0.05}$As. A
trench is etched in the GaMnAs layer, resulting in a SWCNT contacted
by two GaMnAs contacts (see figure \ref{Jensen}). The Curie
temperature of the GaMnAs layer is $\sim$ 70 K. In electron
transport measurements Coulomb blockade is observed at low
temperatures, as well as indications for TLL behaviour
\cite{2004-jensen-nanolett}.

In a later publication \cite{2005-jensen-PRB}, Jensen \textit{et
al}$.$ discuss MR measurements on these devices and compare them
to devices with metallic FM contacts. At $T$ = 0.3 K, a
reproducible MR sign alteration is found in a small $V_g$-range.
High MR ratios on the order of 100\% have been reported. Somewhat
troublesome, hysteretic MR was also found for devices with only
one FM contact. Jensen \textit{et al}$.$ cannot provide an
explanation for the large MR, the sign change and the fact that MR
also shows up with one FM contact.


\begin{figure}[htbp]
\center{\includegraphics[width=8cm]{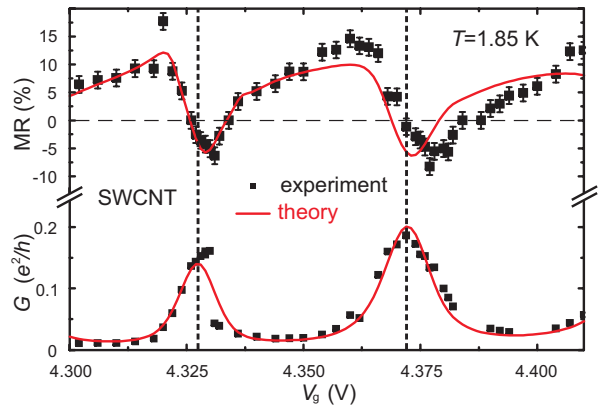}}
  \caption{MR and linear conductance as a value
  of gate voltage for a SWCNT at 1.85 K. The separation between the contacts
  is 500 nm. The red line gives a fit with the theory of the spin-dependent
  Breit-Wigner transmission probability. Courtesy of S. Sahoo\cite{2005-sahoo-natphys}.}
  \label{SahooSWCNT}
\end{figure}

The MWCNT measurements of Sahoo \textit{et al}$.$, discussed in
section \ref{MWCNT exp}, were extended in the same study by
measurements on SWCNTs. The results on SWCNTs were used to
substantiate that interference of single-particle levels is the
physical origin for the observed MR oscillation
\cite{2005-sahoo-natphys}. QD behaviour is already observed at 1.85
K in the SWCNTs (as compared to 0.3 K for the MWCNTs, see
figure\ref{SahooMWCNT}), as both the charging energy and the energy
level spacing is larger for SWCNTs. The MR changes sign on each
conductance resonance (see figure \ref{SahooSWCNT}) and varies
between -7\% and 17\%. The higher MR values as compared to the
MWCNTs are ascribed to the larger charging energy in SWCNTs
\cite{2000-Barnas-PRB}. An order of magnitude lower MR is found for
NM-SWCNT-NM control devices.

Sahoo \textit{et al}$.$ explain the systematic change of MR sign
with gate voltage by combining the spin-dependent Breit-Wigner
formula for resonant tunnelling through a QD coupled to FM
electrodes \cite{Brataas2000,Waintal2000,Braun2004} with the
Julli\`{e}re TMR expression \cite{1975-julliere-physlett}. It is
shown that on resonance the MR is negative, whereas off-resonance it
is positive (see figure \ref{SahooSWCNT}). For reproducing the
observed asymmetry in the MR, it is assumed that the QD energy
levels are spin-dependent, resulting in reasonable agreement between
experiment and theory. A quantitative theory based on the
Landauer-B\"{u}ttiker model describing the MR data, was presented in
\cite{Cottet2006}.


As opposed to the Coulomb blockade regime described above, Man and
coworkers \cite{Man} have investigated a SWCNT connected to
\emph{highly transparent} PdNi contacts. In this regime, the SWCNT
device acts like a Fabry-P\'{e}rot (quantum) interferometer, and not
as a QD as in the case of \cite{2005-sahoo-natphys}. The observed MR
with a maximum of 4 \% at 4.2 K oscillates with gate voltage in
phase with the resistance, which excludes magneto-Coulomb effects
and can be fitted with a model based on the Landauer-B\"{u}ttiker
picture. Oscillations as a function the bias voltage are also seen,
and can be understood in terms of the model. Decrease of the MR with
increasing bias is attributed to the decrease in polarization of the
electrodes when the bias voltage approaches the exchange energy of
PdNi.


As argued in section \ref{spurious}, there are a number of
spurious effects that can obscure the MR due to the spin-valve
effect. In order to eliminate these spurious MR effects, Tombros
\textit{et al.} for the first time used a 4-terminal, non-local
geometry to measure spin accumulation in SWCNTs
\cite{2006-tombros-prb}. The non-local geometry (see figure
\ref{nonlocal}) allows to separate the spin and charge currents
and was successfully applied to inorganic (metallic) systems
before \cite{Jedema,JohnsonSilsbee,2002-Jedema-Nature}.
Importantly, Tombros \textit{et al}$.$ conclude that the MR in
conventional, two-terminal measurements (as applied in all
experiments discussed so far) is dominated by spurious effects.

SWCNTs are deposited on a SiO$_2$ substrate, and Co contacts are
fabricated using EB lithography and EB evaporation. In the
non-local geometry, at least two out of the four contacts should
be FM. For practical reasons, Tombros \textit{et al}$.$ use four
FM contacts with different coercive fields. It is essential that
the contacts are low-ohmic (to avoid QD formation), and that
charge and spin transport is possible beneath them. In their best
device, a 4-terminal conductance of 2.5 $e^2/h$ is measured,
indicating that at least one metallic SWCNT is probed (bundles of
a few SWCNT cannot be excluded however). Comparing the 2-terminal
and 4-terminal resistances, contact resistances around a k$\Omega$
are deduced, much lower than in the case of e.g. Tsukagoshi
\textit{et al}$.$
\cite{1999-Tsukagoshi-Nature,1999-Tsukagoshi-PhysicaE}. Comparing
the results of 2-terminal (local) and 4-terminal (non-local)
measurements, it is concluded that 90\% of the MR in the
two-terminal measurement cannot be attributed to spin
accumulation. Spin accumulation hence seems to be easily
overshadowed by spurious effects.


Recently, Alphenaar's group has succeeded in contacting SWCNTs with
Ni electrodes spaced 10 nm apart using shadow evaporation
\cite{2006-Alphenaar-APL}. The comparison of these extremely short
devices with earlier devices may shine more light on the role of
spurious effects.


In addition to CNTs, also spin transport through C${_60}$ films is
studied. Zare-Kolsaraki and Micklitz
\cite{2004-zare-EPJ,2004-zare-MMM,2003-zare-PRB,2004-zare-PRB,2006-zare-MMM}
studied MR in granular films of Co clusters mixed with C$_60$
molecules. MR is observed in films with a Co fraction between 0.23
and 0.32. The highest MR they measure is around 30\% at 4K and
drops to a few percent at 60K. Miwa \emph{et al}$.$
\cite{2006-Miwa-JJAP} studied the same system, but performed
ex-situ measurements (i$.$e$.$ not under vacuum conditions as in
the case of Zare-Kolsaraki and Micklitz), allowing for better
characterization of their device. Good quality devices give a MR
of 8 \% at 4.2K and around 0.1 \% at room temperature.


\subsection{Carbon nanotubes summary}
\label{CNT concl}

\begin{table}
\caption{Carbon nanotube spin valve devices}
\begin{tabular}{lllll}
\hline
  Nanotube & FM contacts & MR = $(R_{AP}-R_P)/R_P$ (\%) & $T$ (K) & ref.  \\
  \hline
  MW & Co/Co & +9 & 4.2 & \cite{1999-Tsukagoshi-Nature} \\
  MW & Co/Co & +2.5, +9 & 4.2 & \cite{2003-Chakraborty-APL} \\
  MW & Co/NiFe & -2.2, +0.6 & 14 & \cite{2001_Orgassa_nanotech} \\
  MW & Co/Co & +30, -36 & 4.2 & \cite{2002-schneider-APL,2002-schneider-JAP} \\
  MW & PdNi/PdNi & oscillation from -5 to +6 & 1.85 & \cite{2005-sahoo-natphys} \\
  MW & LSMO/LSMO & +40 & 4.2 & \cite{2006-Hueso-APL} \\
  MW & LSMO/LSMO & +61 & 5 & \cite{2007-hueso-Nature} \\
  MW & Co/Co & 3.6 & 2.5 & \cite{2004-hoffer-EPL} \\
  SW & PdNi/PdNi & oscillation from -7 to +17 & 1.85 & \cite{2005-sahoo-natphys} \\
  SW & LSMO/LSMO & +61 & 5 & \cite{2007-hueso-Nature} \\
  SW & Co/Co & +2.6 & 2.0 & \cite{2002-kim-PRB} \\
  SW & Fe/Fe & +100 & 0.3 & \cite{2003-Jensen-worldscientific} \\
  SW & (Ga,Mn)As/(Ga,Mn)As & 75 & 0.3 & \cite{2005-jensen-PRB} \\
  SW & PdNi/PdNi & +4 & 4.2 & \cite{Man} \\
  SW & Co/Co & +6 & 4.2 & \cite{2006-tombros-prb} \\
  SW & Ni/Ni & oscillation from -6 to +10 & 4.2 & \cite{2006-Alphenaar-APL} \\
\hline
\end{tabular}
\label{CNTtable}
\end{table}

An overview of spintronic experiments on CNTs is given in Table
\ref{CNTtable}. So far, MR is only observed at low temperatures with
a maximum of 120 K in \cite{2007-hueso-Nature}. Metals (Co, NiFe,
PdNi), half-metallic oxides (LSMO) and magnetic semiconductors
(GaMnAs) have been used as FM contacts. Reasonably long spin
relaxation times (up to 30 ns) and lengths (1.4 - 50 $\mu$m) have
been derived. However, these numbers are not based on the MR signal
for a reasonable amount of different contact separations (allowing
for an exponential fit as explained in section \ref{CPP GMR}).
Instead, they are based on only one contact separation and (in most
cases) the simple Julli\`{e}re model, resulting most probably in
underestimation of the spin relaxation time.

Although spin injection and detection in CNTs seems feasible, the
large variation in obtained results is striking. Probably one of the
major issues, if not \emph{the} major issue, is the quality of the
CNT-contact interface. Krompiewski \cite{Krompiewski1} has shown
that the GMR in a CNT depends on the detailed properties of the
interface. Especially the introduction of an oxide layer in between
the metal and the CNT, of which the coupling is assumed to be
antiferromagnetic (as in the case of CoO, but also for NiO and FeO),
can drastically influence the GMR ratio. The GMR signal also depends
on the position of the chemical potential in the contacts.

For the case of MWCNTs, it is shown \cite{Krompiewski2} that the GMR
signal around $E_F$ strongly depends on the interaction between
outer and inner tubes (even when the current solely flows through
the outer shell). This may be one of the reasons for the observed
negative GMR in some experiments. The degree of disorder in the CNT
(impurities, dopants or incommensurate inner shells) has been
identified as a cause of GMR reduction \cite{Krompiewski3}.

The gate-voltage dependence of the MR in CNTs has been studied for
both weak coupling to the contacts (QD behaviour), as for
transparent contacts (quantum interference behaviour). Some of these
results can be explained in terms of the Landauer-B\"{u}ttiker
formalism, but more experimental work is needed to convincingly
exclude spurious MR effects. The application of the non-local
geometry of \cite{2006-tombros-prb} may be crucial in this respect.


\section{Experiments on organic thin-film, SAM and single-molecule devices}
\label{OTF exp}

In this section, we discuss experiments on organic spin valves where
the spacer between the FM electrodes is formed by an organic thin
film. Thin films of polymers or small molecules are easy to
fabricate, and are already applied in different technological
applications such as OLEDs (see section \ref{OTF}). The carrier
mobility in organic thin films however, is much lower than in CNTs.
For measuring any sizeable MR effect in a SV structure, it is
therefore necessary to have a small ($\sim$100 nm) spacing between
the FM contacts.

\subsection{Organic thin-film spin valves}


An example of an organic semiconductor that is broadly applied in
OLEDs is sexithienyl (T$_6$), a $\pi$-conjugated rigid-rod oligomer.
Its HOMO level is 4.8 eV \cite{2000-Hill-ChemPhysLett}, and
depending on its morphology, the mobility ranges from 10$^{-2}$ to
10$^{-4}$ cm$^2$V$^{-1}$s$^{-1}$
\cite{1998-Torsi-PRB,1992-Ostoja-AMOE}. Dediu \textit{et al.}
\cite{2002-Dediu-SSC} applied this p-type material in the first
report on spin injection in an organic semiconductor. Importantly, a
MR effect was measured at \emph{room temperature}.
\begin{figure}[hbtp]
  \center{\includegraphics[width=8cm]{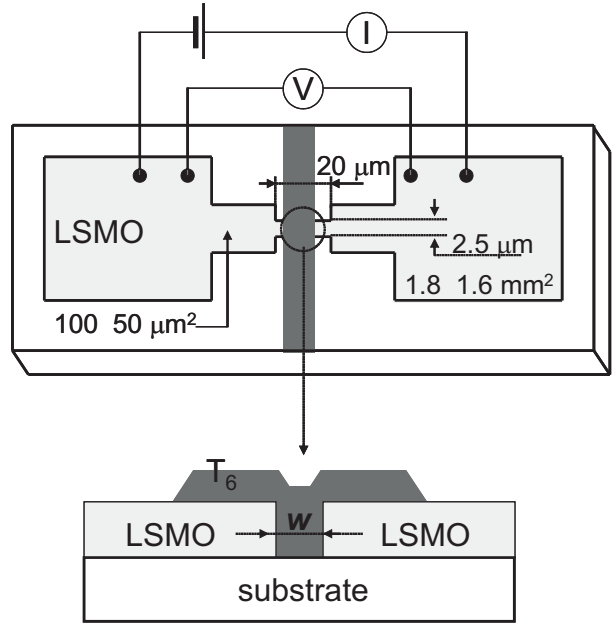}\\}
  \caption{Schematic top view and cross section of a hybrid
  LSMO/T$_6$/LSMO junction. An epitaxial thin film of LSMO was
  deposited on matching substrates (NdGaO$_3$, SrTiO$_3$), and electrodes
  were fabricated by EB lithography.
  The separation $w$ of the electrodes varied between 70 and 500
  nm on one single substrate. T$_6$ films (100 - 150 nm thick)
  were deposited on top of the electrodes by molecular beam deposition.
  The resistance of the films was typically 10$^2$-10$^3$ for
  thicknesses of 100 nm. Courtesy of V. Dediu\cite{2002-Dediu-SSC}.}
  \label{dediu}
\end{figure}
EB lithography is used to fabricate electrodes out of a thin
($\sim$100 nm) film of La$_r$Sr$_{1-r}$MnO$_3$ ($0.2 < r < 0.5$)
deposited epitaxially on an insulating substrate, see figure
\ref{dediu}. LSMO is a half-metal with ideally 100\% spin-polarized
carriers at the Fermi level below the Curie temperature ($T_C \sim
370K$). Unlike FM contacts as Co, Ni or Fe, LSMO is stable against
oxidation. The thin (100-150 nm) T$_6$ films are deposited on the
substrate by molecular beam evaporation. The workfunction of LSMO is
estimated to be around 5 eV, close to the HOMO level of T$_6$. The
observed linear $I-V$ characteristics seem to confirm the low-ohmic
contact between LSMO and T$_6$. If the spin polarization in the LSMO
contacts is indeed (close to) 100\%, the conductivity mismatch
problem (see section \ref{mismatch}) should not play a role and the
low-ohmic LSMO-T$_6$ contact should not prevent spin injection.

As the geometry (and hence the coercive field) of the LSMO
electrodes is the same, Dediu \textit{et al.} do not succeed in
switching the magnetization of each FM contact independently in a
controlled fashion. However, they change the relative orientation
from random, at low field, to parallel at higher field. A maximum
resistance decrease of $\sim$30\% from the random to the parallel
configuration is observed for a 140 nm channel at room temperature.
The MR is independent of field orientation (perpendicular or
parallel), and no MR effect is observed for channels larger than 200
nm. The contact geometry complicates the evaluation of the spin
relaxation length. A rough estimate of $l_s$ = 200 nm is made. Using
this value and a mobility of 10$^{-4}$ cm$^2$V$^{-1}$s$^{-1}$, one
finds a spin relaxation time $\tau_s \sim$ 1 $\mu$s.


\begin{figure}
  \center{  \includegraphics[width=8cm]{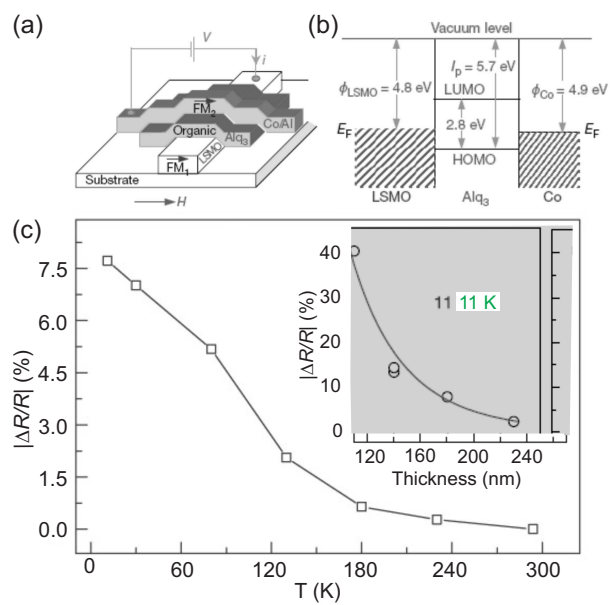}}
  \caption{(a) Schematic picture of an organic SV device. For the
  bottom electrode (FM$_1$) a LSMO film is used, on which the
  organic semiconductor Alq$_3$ is thermally evaporated. A thin Co
  film (3-6 nm), serving as the second FM contacts (FM$_2$), and Al
  contact are then evaporated using a shadow mask without breaking
  the vacuum, resulting in an active device area of 2 x 3 mm$^2$. A
  magnetic field is applied parallel to the substrate as denoted in
  the figure. (b) Schematic band diagram of this device in the rigid
  band approximation. The Fermi levels ($E_F$) and the workfunctions
  ($\phi$) of LSMO and Co are shown, as well as the HOMO-LUMO levels
  of Alq$_3$. (c) MR as a function of temperature
  measured at a voltage of 2.5 mV for an Alq$_3$ thickness of 160 nm.
  The inset shows the MR as a function of thickness at a temperature
  of 11 K. The line through the data point is a three parameter fit
  to an adjusted Julli\`{e}re model. Courtesy of Z. H. Xiong\cite{2004-xiong-nature}.}
  \label{Xiongfig}
\end{figure}

Xiong \textit{et al.}\cite{2004-xiong-nature, 2005-xiong-IEE} used
the small $\pi$-conjugated molecule 8-hydroxy-quinoline aluminium
[Alq$_3$, see figure \ref{OTFT}(b)] as a spacer in a (vertical) SV
device, see figure \ref{Xiongfig}(a). Alq$_3$ is a popular material
for use in OLEDs, as it is easily deposited and can be combined with
various metallic electrodes. LSMO is used as bottom contact ($H_c
\sim$30 Oe), and Co ($H_c \sim$ 150 Oe) as the top electrode [see
figure \ref{Xiongfig}(a)]. Xiong \textit{et al.} thereby succeeded
in realizing both P and AP magnetization of the contacts. A
schematic band diagram is given in figure \ref{Xiongfig}(b),
indicating the HOMO and LUMO levels of Alq$_3$ and the (nearly
equal) workfunctions of LSMO and Co. At low bias voltages, holes are
injected from the anode into the HOMO level. The evaporation of the
top Co electrode causes pinholes and Co inclusions in the Alq$_3$
layer over a distance of $\sim$100 nm. The Co/Alq$_3$ interface is
therefore poorly defined.

A \textit{negative} MR of 40\% is observed at 11 K for a 130 nm
thick Alq$_3$ layer. Inverse MR has also been reported for
LSMO/SrTiO$_3$/Co and LSMO/Ce$_{0.69}$La$_{0.31}$O$_{1.845}$/Co
MTJs, and is ascribed to negative spin polarization of the Co
d-band. The MR strongly depends on bias voltage and temperature, and
disappears for $|V| \gtrsim$ 1 V, or $T \gtrsim$ 300 K. The
temperature dependence [figure \ref{Xiongfig}(c)] is dominated by
spin relaxation in Alq$_3$ (and not by the temperature dependence of
the magnetization of the LSMO), as confirmed in SV devices where
LSMO is replaced by Fe \cite{2005-xiong-synmet} and by
photoluminescence measurements.

It is hard to make a good estimate for $l_s$ in this geometry, but a
value of $\sim$45 nm at 11 K is reported. The ill-defined Co/Alq$_3$
interface is somewhat troublesome in this experiment, and makes it
difficult to interpret the results. As discussed in Part I, the role
of the interfaces is extremely important in spintronic devices. Next
to the low-$T$ MR that is ascribed to the SV effect, a strong
negative MR is observed for higher magnetic fields ($H$ = $\sim$1-10
kOe) at higher temperature. The latter MR effect is referred to by
the authors as \emph{high-field magnetoresistance}, or HFMR. In a
later work \cite{2004-xiong-prl} it is argued that this HFMR can be
related to the LSMO electrode (no HFMR is found in devices without
LSMO contacts). Although single LSMO films already show HFMR, the
effect is enhanced by orders of magnitude when the LSMO forms an
interface with an organic semiconductor. For the proposed mechanism
we refer to \cite{2004-xiong-prl} and references therein.

Dediu \textit{et al.} find a considerable MR at room temperature
(over a magnetic field scale of several kOe) in their
LSMO/T$_6$/LSMO, whereas Xiong \textit{et al.} only find HFMR at
higher temperatures (which is not related to the SV effect). The
observed MR by Dediu and coworkers is therefore possibly also
related to this HFMR. Distinction between GMR and HFMR could be
made, by introducing asymmetry between the LSMO contacts in
\cite{2002-Dediu-SSC}. Room-temperature SV behaviour was also
recently claimed in LSMO/P3HT/Co \cite{2006-Majumdar-APL} and
LSMO/P3OT/LSMO \cite{2006-Kumar-SSC} devices, where P3HT is
poly(3-hexyl thiophene) and P3OT is poly(3-octylthiophene), both
semiconducting $\pi$-conjugated polymers.


Santos \textit{et al}$.$ \cite{2007-santos-PRL} demonstrate
spin-polarized \emph{tunnelling} through a thin Alq$_3$ barrier
sandwiched between a Co (bottom) and Ni$_{80}$Fe$_{20}$ (permalloy,
Py) contact (top) at \emph{room temperature}. $I-V$- and
polarization measurements indicate the good quality of the Alq$_3$
barrier without any cobalt inclusions. The TMR value is improved by
adding an Al$_2$O$_3$ layer in between the Co and the Alq$_3$ tunnel
barrier, which reduces the formation of interfacial charge states.
The highest TMR they find at room temperature is 6 \% and a
substantial TMR value is even present above 100 mV. They measure a
positive polarization for Co and Py, which corresponds to the
observed positive TMR, but is in contrast to the negative MR
reported by Xiong \textit{et al}$.$ \cite{2004-xiong-nature}. Santos
\textit{et al}$.$ argue that this is not because of the negative
polarization of the Co d band, as proposed by Xiong \textit{et
al}$.$, but might originate from the opposite spin asymmetry
coefficients of Co and LSMO. They also state the role of the Co
inclusions is unclear.


Riminucci et al$.$ \cite{2007-Riminucci-Arxiv} report on MR
measurements on a vertical LSMO/Alq$_3$/tunnel barrier/Co device.
The tunnel barrier is SiO$_2$ or LiF, and in both cases a negative
MR of a few percent was detected up to 210 K. They especially
focus on the interface between LSMO and Alq$_3$ in order to
investigate the character of the charge carriers in Alq$_3$.
Alq$_3$ is an n-type material, whereas the unperturbed energy
levels suggest hole conduction. Photo-electron spectroscopy
reveals a strong interface dipole, which shifts the energy levels
of the system. They construct a semi-quantitative energy level
model which predicts electron transport due to a smaller energy
barrier between the Fermi energy of LSMO and the LUMO of Alq$_3$
than between the Fermi energy and the HOMO level. The model also
indicates a resonance between the LUMO level and the Co d-band,
which provides an explanation for the negative SV effect.



The only experimental work so far that addresses the question
which spin relaxation mechanism (see section \ref{spinrelax}) is
dominant in organic semiconductors, is the study by Pramanik
\textit{et al}$.$ \cite{Pramanik}. It was argued that 1D
confinement suppresses the DP mechanism
\cite{2004-Pramanik-APL,2005-Pramanik-IEEE}, whereas it enhances
the EY mechanism. The effect of 1D confinement on the spin
relaxation time should therefore provide an answer to the question
which relaxation mechanism is dominant. Pramanik \textit{et al}$.$
studied ensembles of Co/Alq$_3$/Ni nanowires and observed a MR
effect at low temperature. By comparing their results with those
of Xiong and coworkers \cite{2004-xiong-nature}, they conclude
that the dominant spin relaxation mechanism in organics is the EY
mechanism. More studies, also in different organic systems, will
be necessary to tell whether this conclusion is generally valid.


\subsection{Organic magnetoresistance (OMAR)}

Recently, there have been a number of studies -- mainly by Mermer
and Wohlgenannt and coworkers -- reporting a considerable MR
effect in organic semiconductor devices \emph{without} FM
contacts, referred to as ``organic magnetoresistance" or OMAR
\cite{2006-prigodin-synmet,2005-wohlgenannt-PRB,2006-Kumar-SSC,2004-wohlgenannt-NJP,2005-wohlgenannt-SSC,2006-wohlgenannt-PRB,2006-wohlgenannt-archiv}.
In experiments on the polymer polyfluorene (PFO)
\cite{2004-wohlgenannt-NJP}, the small molecule Alq$_3$
\cite{2005-wohlgenannt-SSC} and several more $\pi$-conjugated
polymers and small molecules \cite{2005-wohlgenannt-PRB}, it was
shown that this OMAR (defined as $\Delta R/R = (R(B)-R(0)/R(0))$)
is quite universal in nature, can be either positive or negative,
and reaches values up to 10\% for $B \sim$ 10 mT at room
temperature. OMAR is shown to be related to the bulk resistance of
the organic film \cite{2004-wohlgenannt-NJP,2005-wohlgenannt-SSC}.
Depending on the organic material, OMAR obeys the empirical law
$\Delta R(B)/R \propto B^2/(B^2 + B_0^2)$ or $\Delta R(B)/R
\propto B^2/(|B| + B_0)^2$, where $B_0 \approx$ 5 mT in most
materials \cite{2005-wohlgenannt-PRB}, increasing with spin-orbit
coupling. The effect is only weakly dependent on temperature, is
independent of magnetic field direction and impurities, and it
typically decreases with increasing voltage and carrier density
\cite{2004-wohlgenannt-NJP,2005-wohlgenannt-SSC}. It was argued
that OMAR is a bulk effect related to the majority carrier current
only (holes in PFO and electrons in Alq$_3$). In
\cite{2006-wohlgenannt-PRB} hyperfine interaction (see section
\ref{hyperfine}) was explored as a possible cause of OMAR.
Although the model presented could explain some characteristics of
OMAR, open questions remained.

Prigodin \textit{et al}$.$ \cite{2006-prigodin-synmet} recently
proposed a model to explain OMAR based on the assumption that charge
transport in organic semiconductors is electron-hole recombination
limited. It is argued that in the space-charge-limited transport
regime (i$.$e$.$ there are no free carriers induced by a gate as in
an OTFT, see section \ref{OTFT}), both electrons and holes are
injected (possibly with very different mobilities). The electron and
holes can form electron-hole (e-h) pairs, that are either in the
singlet (S) or triplet (T) state. It is shown that the
space-charge-limited current density increases with decreasing e-h
recombination rate. As the recombination rate depends on the degree
of mixing between S and T states, the recombination rate -- and
hence the current density -- becomes $B$-dependent. However, the
experimental fact that OMAR is weakly dependent on the minority
carrier density (and also occurs in heavily p-doped devices) is not
in agreement with the Prigodin model.

Another model has been put forward by Wohlgenannt
\cite{2006-wohlgenannt-archiv}. He explains OMAR by hopping
accompanied by spin-flip processes. Hopping from a single occupied
(SO) state to another SO state, forming a double occupied (DO)
state, is normally not allowed when the spins are parallel, due to
the Pauli principle (only singlet states are allowed on a DO
state). Spin-flip mechanisms like the hyperfine interaction
however, make this transition possible. Because the hyperfine
interaction in organic materials is very small (see section
\ref{hyperfine}), the spin-dynamics is suppressed by applying a
small magnetic field of typically several mT (the associated
Zeeman energy is large enough to pin the spins), resulting in a
positive MR. A model incorporating the spin-flip probability is in
agreement with several experiments, but can not explain the
observed negative MR. A possible mechanism for negative MR is a
reduction of the formation of bipolarons, which are believed to
have a reduced smaller hopping probability than polarons due to
their larger lattice distortion and associated larger mass. In
this case, applying a magnetic field leads to less DO states and
therefore to less bipolarons, resulting in a larger hopping
conductance.


\subsection{Single-molecule devices and self-assembled monolayers}

Although the organic semiconductor devices described above have very
thin layers polymers or small molecules, these layers are still
typically thicker than $\sim$100 nm. In the search for
miniaturization of electronic functional devices, molecular
monolayers, and eventually single molecules are the ultimate limit
(see section \ref{singlemolecules}). The spintronic properties of
such systems are not explored extensively yet, but a few interesting
studies are discussed below.


Using time-resolved Faraday rotation spectroscopy, Ouyang and
Awschalom demonstrated coherent spin transfer of photo-excited
carriers between semiconductor QDs through conjugated molecules at
room temperature \cite{2003-Ouyang-Science}. Their devices consist
of multilayer CdSe QDs solids that are bridged by
1,4-benzenedimethanethiol, and they show that the spin transfer
efficiency is $\sim$20\%.


Pasupathy \emph{et al}$.$ \cite{2004-pasupathy-science} have
observed the Kondo effect in single C$_{60}$ molecules connected to
two Ni electrodes. The electrodes are made by electro-migration
after which a solution of C$_{60}$ is dispersed onto them. Different
shapes of the contacts allow for independent switching of their
magnetization. The Kondo behaviour is confirmed by the dependence of
the conductance on the temperature and the magnetic field. TMR
values of -38\% and -80\% are found, much larger than the predicted
value of 21\% by the Julli\`{e}re model, which is explained by the
fact that for AP magnetization, the Kondo resonance occurs closer to
the Fermi energy, thereby enhancing its conductance.


The first transport measurements on a SV involving a single
molecular layer were reported by Petta \textit{et al}$.$
\cite{2004-Petta-PRL}. The device is a nanometer-scale magnetic
tunnel junction in the nanopore geometry
\cite{1999-Chen-Science,1989-Ralls-APL}, consisting of a SAM of
octanedithiol (100-400 molecules) sandwiched in between Ni
contacts. The transport properties of octanethiol with NM contacts
have been extensively studied
\cite{1999-Bumm-JPhysChem,2001-Wold-JACS,2002-Cui-Nanotech,2003-Wang-PRB}.

The majority of the octanedithiol devices have resistances larger
than $h/e^2$, implying tunnelling transport. A tunnelling barrier
height of 1.5 eV is found in reasonable agreement with earlier work
on octanethiols on Au \cite{2003-Wang-PRB}. Both positive and
negative MR up to 16\% is reported for low bias voltage at 4.2K,
rapidly decreasing with bias voltage and temperature. The largest MR
is measured for the most resistive devices. The Julli\`{e}re formula
(using the definition TMR = $R_{max}-R_{min}/R_{min}$) is TMR =
$2P_1P_2/(1-P_1P_2)$. Using $P_1=P_2=0.31$ for Ni, one finds TMR =
21\%, somewhat larger than the experimentally observed MR values.
Several test devices are made in order to rule out artifacts of the
fabrication process. As mentioned in section \ref{TMR}, localized
states in the SAM tunnel barrier could possibly explain the
anomalous behaviour. Also the observed telegraph noise may be due to
imperfections in the molecular barrier.

In \cite{Rocha} Rocha and coworkers calculated the transport
properties through two different molecules sandwiched between two
ferromagnetic leads. They predict both TMR and GMR behaviour for
octane and 1,4-[n]-phenyl-dithiolate, respectively.


One step further than the above work of Petta \textit{et al.} would
be to integrate the spintronic functionality \emph{within} the
molecule. Liu \textit{et al.} \cite{2005-Liu-Nanolett} have proposed
a single-molecule spin switch and SV based on the organometallic
molecule dicobaltocene between NM contacts. The singlet (AP) spin
state blocks electron transport near the Fermi energy, while the
triplet (P) state enables much higher current. The S-T energy
splitting depends on the insulating spacer between the cobaltocenes.

\subsection{Organic thin-film, SAM and single-molecule devices summary}

\begin{table}[htbp]
\caption{Organic thin-film, SAM and single-molecule devices}
\begin{tabular}{lllll}
\hline
  Material & FM contacts & MR = $(R_{AP}-R_P)/R_P$ (\%) & $T$ (K) & ref.  \\
  \hline
  T$_6$                     & LSMO/LSMO         & +30               & 300   & \cite{2002-Dediu-SSC} \\ 
  Alq$_3$                   & LSMO/Co           & 40                & 11    & \cite{2004-xiong-nature} \\ 
  P3HT                      & LSMO/Co           & 80/1.5            & 5/300 & \cite{2006-Majumdar-APL} \\
  Alq$_3$ (tunnel barrier)  & Co-Al$_2$O$_3$/Py & +6                & 300   & \cite{2007-santos-PRL} \\
  Alq$_3$ (tunnel barrier)  & LSMO/SiO$_2$,LiF-Co         & -2.5              & 100   & \cite{2007-Riminucci-Arxiv} \\
  Octanedithiol             & Ni/Ni             & -6 to +16         & 4.2   & \cite{2004-Petta-PRL} \\
\hline
\end{tabular}
\label{comparison}
\end{table}

MR has been reported in a few organic semiconductors at low
temperatures. Although the mobilities in these materials are rather
low, reasonably large spin relaxation lengths (100 – 200 nm) and
accompanying large spin relaxation times (1 $\mu$s) have been
reported. These values are however still rough estimates. The
experiments seem promising for spin injection into organic
semiconductors, but some remarks have to be made. In the case of the
early T$_6$ experiment \cite{2002-Dediu-SSC}, the symmetry of the
two contacts makes the interpretation of the data difficult. HFMR
has also been suggested to be the origin of the observed effect. For
the Alq$_3$ experiments, EB evaporation of the top contacts most
probably introduces impurities in the material. Non-destructive
contacting is need for reliably extracting reliable values for the
spin relaxation length and time.

A few experimental studies exist on spin transport through a SAM or
single molecule. These experiments are very interesting from a
fundamental point of view and could provide more information about
tunnelling and transport mechanisms in organic spintronic devices.

\section{Conclusions}
\label{Concl}

Since the first reported organic spintronic device in 1998,
devices with a variety of organic spacer materials have been
experimentally investigated. Most research so far has been devoted
to CNTs. They offer low spin-orbit coupling and hyperfine
interaction (and hence expectedly long spin relaxation times) in
combination with high carrier mobilities. In addition, CNTs allow
for studying the interplay of spin transport with Coulomb charging
and quantum interference effects. Organic thin films have been
investigated and there are a few studies on spin-dependent
tunnelling through SAMs and single-molecule structures. Thin-film
devices are receiving an increasing amount of attention, not in
the last place because of the recent discovery of OMAR in organic
semiconductors. The recently discovered 2D form of carbon,
graphene, is very promising for spintronic research as well as for
the realization of spin-based quantum bits.

Although a lot of promising results have been obtained in the last
decade, there are still a number of serious problems to be tackled.
Control and understanding of the ferromagnet-organic interface is
one of the major issues. Because of the fragile nature of organic
materials, conventional microfabrication techniques cannot be used
without caution. Non-destructive contacting of organic materials is
important in organic electronics in general, but of crucial
importance in organic spintronics in particular. A lot of effort is
needed in this direction. Besides reliable contacts, also
well-defined and clean organic spacer materials are needed. Defects
and impurities in the organic materials themselves is certainly also
a cause of the large scatter in the observed MR values. Even in the
absence of defects or impurities, the combination of different
materials can lead to new behaviour, for example due to the
formation of an interface dipole layer.

Most spin relaxation lengths reported so far for organic devices
are comparable to metals and inorganic semiconductors. The largest
relaxation length \cite{2007-hueso-Nature} is however an order of
magnitude larger than electrically detected in GaAs (6 $\mu$m)
\cite{Lou} and almost two others of magnitude bigger than that
electrically measured for metals (around 1 $\mu$m at 4.2 K and
several hundreds of nm at room temperature \cite{Jedema}, which
seems very promising for organic spintronics when for example
integrated in a FET structure. However, these values have to be
considered with a little care. Most of the time, the relaxation
length is obtained with a modified Julli\`{e}re model for only one
contact spacing and does therefore not account for spin relaxation
at the interface. A more elaborate model, like the one provided by
Jaffr\`{e}s and Fert, could lead to more accurate calculations.
Distance-dependent studies, in which the injector-detector is
varied within the same device, are necessary. Another problem is
separating true SV signals from spurious effects. These effects
can be ruled out with non-local measurement.

The study of spin injection, transport, and relaxation in organic
materials is still in its infancy. There is still a lot to find out
about the role of the transport mechanism (band or hopping
conduction, polaronic transport) on spin transport. Especially for
finding the relevant spin relaxation mechanisms in organic
materials, more theoretical and experimental work is needed.


The discovery of TMR and GMR in metallic spin valves has led to a
revolution in magnetic memory. Parallel to this, the application of
organic materials for electronic devices meant another revolution.
The merging of spintronics and organic electronics is likely to
ultimately lead to new spin-based, versatile devices and possibly
even to robust quantum bits for quantum information and computation.
Along the way, certainly many fundamental issues need to be
addressed, being a challenge for scientists from different
disciplines. Organic spintronics is an area of research where
physics, chemistry and electrical engineering inevitably meet, and
hence an intriguing interdisciplinary field of research.

\section*{Acknowledgments}

We thank R. Coehoorn, A. Dediu, R. Jansen, L. Kouwenhoven, C.
Sch\"{o}nenberger, K. Tsukagoshi, L. Vandersypen, and B. van Wees
for fruitful discussions. This work is part of the VIDI research
program `Organic materials for spintronic devices', financially
supported by the Netherlands Organisation for Scientific Research
(NWO) and the Technology Foundation STW.


\end{document}